\documentclass[%
 reprint,
superscriptaddress,
 amsmath,amssymb,
pra,
floatfix,
]{revtex4-2}

\usepackage{graphicx}
\usepackage{dcolumn}
\usepackage{bm}

\usepackage{color,soul}
\usepackage[acronym]{glossaries}
\newacronym{qrw}{QRW}{Quantum Random Walk}
\newacronym{crw}{CRW}{Classical Random Walk}
\newacronym{rw}{RW}{Random Walks}
\newacronym{pic}{PIC}{Photonic Integrated Circuit}
\newacronym{rr}{PRR}{Photonic Ring Resonator}
\newacronym{rms}{RMS}{Root Mean Square}
\newacronym{fsr}{FSR}{Free Spectral Range}
\newacronym{pr}{PR}{photoresist}
\newacronym{fdtd}{FDTD}{Finite-Difference Time-Domain}
\newacronym{fdfd}{FDFD}{Finite-Difference Frequency-Domain}
\newacronym{dlw}{DLW}{Direct Laser Writing}
\newacronym{gui}{GUI}{Graphical User Interface}

\newcommand{\micron}{$\mathrm{\mu}$m}
\newcommand{\tsb}[1]{\textsubscript{#1}}

\newcommand{\eg}[1]{{\color{blue} #1}}
\definecolor{dgreen}{rgb}{0, 0.6, 0}

\begin{document}

\preprint{APS/123-QED}
\title{
Quantum random walks in coupled photonic ring resonators}%

\author{Ricardo M. R. Adão}
\affiliation{%
    INL - International Iberian Nanotechnology Laboratory
    Av. Mestre José Veiga s/n, 4715-330 Braga, Portugal
}
\author{Manuel Caño-García}%
\altaffiliation[Present address: ]{%
    CEMDATIC, ETSI Telecomunicación, Universidad Politécnica de Madrid, Av. Complutense 30, 28040 Madrid, Spain
}
\affiliation{%
    INL - International Iberian Nanotechnology Laboratory
    Av. Mestre José Veiga s/n, 4715-330 Braga, Portugal
}

\author{Jana B. Nieder}%
\affiliation{%
     INL - International Iberian Nanotechnology Laboratory
     Av. Mestre José Veiga s/n, 4715-330 Braga, Portugal
}
\author{Ernesto F. Galvão}
\email[E-mail: ]{ernesto.galvao@inl.int}
\affiliation{%
    INL - International Iberian Nanotechnology Laboratory
    Av. Mestre José Veiga s/n, 4715-330 Braga, Portugal
}
\affiliation{%
    Instituto de F\'{\i}sica, Universidade Federal Fluminense, Niter\'oi, RJ, 24210-340, Brazil
}


\date{\today}
\begin{abstract}
Quantum random walks use interference for faster state space exploration, which can be used for algorithmic purposes. Photonic technologies provide a natural platform for many recent experimental demonstrations. Here we analyze quantum random walks implemented by coherent light propagation in series-coupled photonic ring resonators. We propose a family of graphs modeling these devices and compare quantum and classical random walks on these structures, calculating steady-state and time-dependent solutions. We obtain conditions for quantum advantage in this setting and show how to recover classical random walks by averaging over quantum phases. Preliminary device feasibility tests are carried out via simulations and experimental results using polymeric directional couplers.
\end{abstract}




\maketitle


\section{\label{sec:Intro}Introduction}

    Quantum random walks describe a walker, often a particle, that explores some state space using coherent wave-like dynamics.
    These non-classical dynamics allow the walker to propagate faster in structures with certain symmetries.
    Quantum walks have been shown to be universal for quantum computation \cite{Childs09}, and may deliver computational speed-ups for many different problems \cite{Childs03, Shenvi03, Childs07, Ambainis07}.
    They have also been proposed as a possible mechanism in energy transfer in biological systems such as photosynthetic molecules \cite{Sension07}.
    
    Since the original theoretical proposal \cite{Aharonov93}, many quantum random walks have been proposed.
    Discrete-time quantum walks can use different coins \cite{Tregenna03}, or be defined without coins via alternative quantization procedures \cite{Szegedy04}, or equivalently, using tesselations of the state space \cite{Portugal16}.
    For a review of both continuous- and discrete-time quantum walks, see \cite{Venegas-Andraca12}.
    
    Quantum random walks have been implemented in various systems: neutral trapped atoms in optical lattices \cite{Karski09}, trapped ions \cite{Schmitz09}, and nuclear magnetic resonance \cite{Du03,Ryan05}.
    Photonic implementations are among the most promising approaches in this regard, using either bulk optics or integrated photonic chips.
    Previously implemented photonic quantum random walks include walks on a 1D linear graph \cite{Perets08, Broome2010,Biggerstaff2016}, on a circle
    \cite{Bian2017,Nejadsattari2019}, and in two-dimensional, cycle-free (tree) graphs
    \cite{Caruso2016,Boada2017,Chen2018,Tang2018,Wang2020}, using either continuous time or discrete time steps. Quantum walks with more than one walking particle have also been implemented using photonics in a way that mimics bosonic or fermionic particle behaviors \cite{Sansoni12}.

    Photonic Ring Resonators (PRRs) are ring-like coupled waveguides first introduced as narrow-band optical filters.
    \acrshort{rr}s have been extensively studied \cite{Rabus2007,Bogaerts2012} and find applications ranging from sensors \cite{Kim2016} to micro-lasers \cite{Stern2017} to fast optoelectronic circuits \cite{Moazeni2017}.
    \acrshort{rr}s can be coupled to other \acrshort{rr}s in complex \acrshort{rr} configurations \cite{Bachman2015} or other photonic elements, becoming one of the most widely used building blocks of today's photonic integrated circuits.
    
    In this work, we model the coherent dynamics of light in coupled \acrshort{rr}s as quantum random walks in these structures.
    We describe a family of graphs that model light propagation along multiple series-coupled \acrshort{rr}s and calculate predictions from classical and quantum propagation dynamics on these graphs.
    The quantum model corresponds to coherent light propagation, whereas the corresponding classical model has the same hopping probability between each pair of coupled waveguides but no coherent effects.
    Comparison between the quantum and classical results enables us to find sufficient conditions for demonstrating quantum advantage in the transport efficiency over these structures.
    We show that the classical model is recovered from the quantum model predictions when we average over the phase acquired when photons go around each ring.
    Besides the theoretical modeling, we also report preliminary experimental feasibility studies regarding the implementation of \acrshort{rr}s using polymeric waveguides in air.
    
    This paper is organized as follows.
    In Section~\ref{sec:qrw}, we describe our model for quantum random walks on series-coupled photonic ring waveguides.
    The classical model calculations are described in section~\ref{sec:CRW}, with the quantum model calculations presented in Section~\ref{sec:QRW} and a comparison between the two in Section~\ref{sec:CRWvsQRW}.
    Section~\ref{sec:Experimental} presents a feasibility study based on different materials and designs, highlighting some of the experimental challenges associated with polymeric devices.
    A discussion of the findings is presented in Section~\ref{sec:discussion}, with some concluding remarks in Section~\ref{sec:conclusion}.

\section{Modeling walks in series-coupled photonic ring resonators} \label{sec:qrw}

\begin{figure}
        \includegraphics[width=8.1cm]{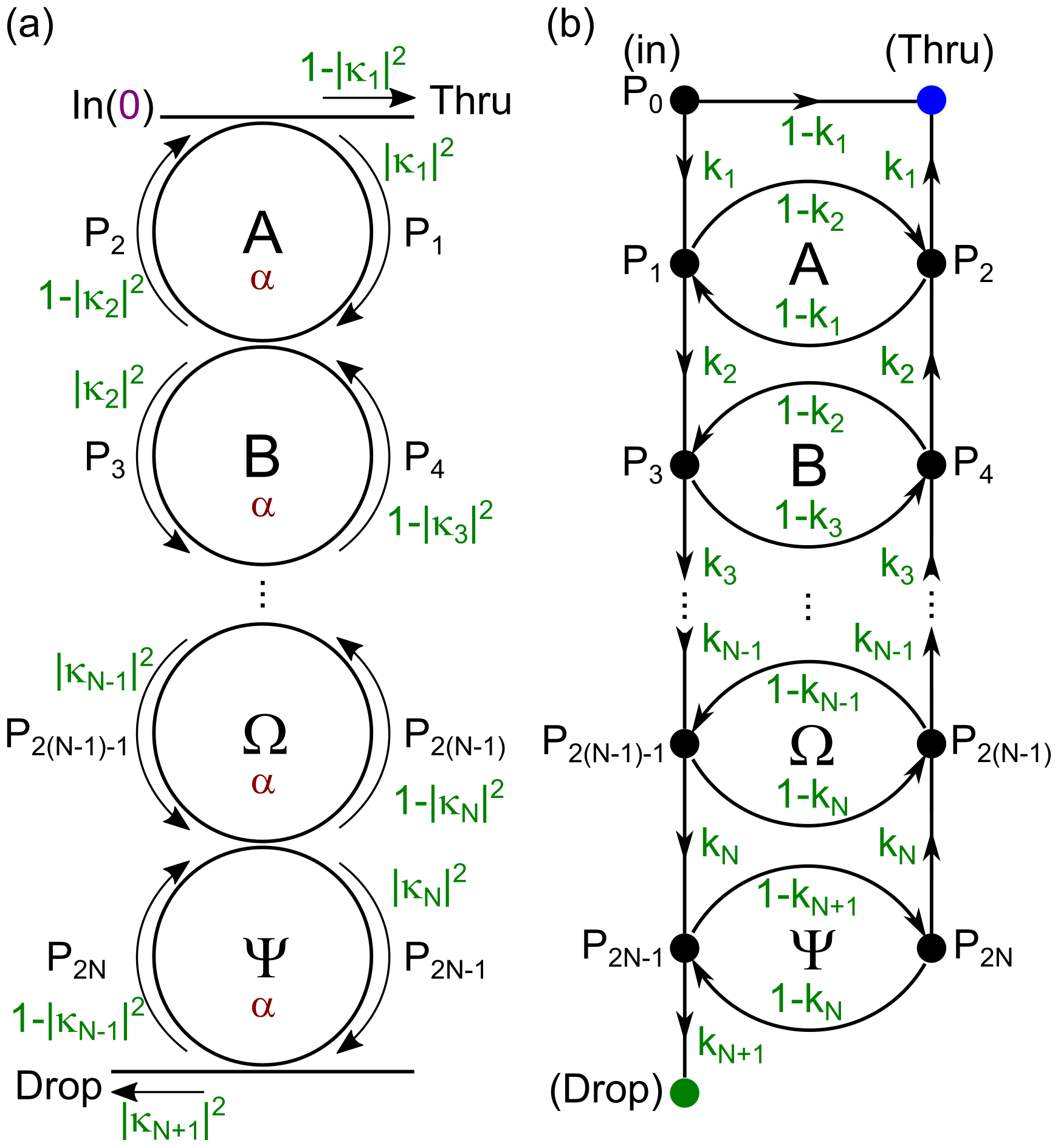}
        \caption{\label{fig:concept}
        Random walks in series-coupled ring resonators.
        (a) Series-coupled multi-ring resonator.
        $\kappa_i$ and $\alpha$ are the coupling coefficient between each indicated pair of waveguides, and loss coefficient, respectively.
        The indicated values correspond to a walker moving consecutively from one ring to the next without looping.
        (b) Graph proposed to model propagation in the series-coupled ring resonators.
        $k_i = |\kappa_i|^2$ are the walker hopping probabilities associated with each pair of nodes.
        The loss coefficient $\alpha$ is omitted from the figure.
    }
    \end{figure}

    In this section, we model light propagation in multiple series-coupled photonic ring resonators, as depicted in Fig.~\ref{fig:concept}(a). 
    The walk begins at the node representing the input port of the system (top-left), where the light enters the device, and ends when the walker reaches either of the output ports: the \textit{Through} node $T$ (abbreviated to \textit{Thru}) or the \textit{Drop} node $D$.
    On the way, the walker hops between half-rings, each of which is represented by an intermediate node $P_i$.
    Let $p(a \to b)$ express the probability of the walker hopping from node $a$ to node $b$.
    The $\kappa_i$ and $\tau_i$ coefficients define the coupling between adjacent waveguides, so that $|\kappa_i|^2+|\tau_i|^2=1$.
    
    Note that the hopping probability depends both on where the walker is and where it is going.
    The physical system composed of coupled photonic ring resonators, as shown in Fig.~\ref{fig:concept}(a), can thus be modeled by the graph in Fig.~\ref{fig:concept}(b), where we identify the input port, the two output ports, all half-rings of the structure, and the hopping probabilities between them.
    The loss coefficient $\alpha$ quantifies the ring round-trip transmission.
    It is omitted from Fig.~\ref{fig:concept}(b) but is considered in the following calculations.
    
    In the proposed photonic implementation, the light injected at the input port propagates continuously through the system, acquiring a phase shift $\theta_i$ around each ring.
    The phase shift depends on the ring radius and wavelength and determines the constructive and destructive interference responsible for the differences between the quantum and classical walks on the same graph structures.
    The physical process of light propagation is continuous in time, but it can be modeled as a discrete process, where each step corresponds to the time it takes for light to propagate around one half-ring.
    This enables analyzing the differences between the quantum and classical solutions in their discrete-time dynamics.
    
    To compare the dynamics of classical and quantum random walks in these structures, we consider a hypothetical puzzle whose goal is to reach the \textit{Drop} port and analyze the parameter-tuning abilities of classical and quantum models to maximize the goal-hitting rate.
    We propose two scalable methods to calculate the probabilities of propagation to the \textit{Thru} and \textit{Drop} ports and, for the sake of simplicity, solve and discuss the simplest single-ring configuration (Fig.~\ref{fig:concept:1ring}).
    However, we present a scalable method that can calculate the probabilities of propagation to any node in the graph for any number of series-coupled rings.
    The double-ring case results are presented in Appendix \ref{ap:CRW:2rings}.

    \subsection{Classical Random Walk}
        \label{sec:CRW}
        
        Let $p_C(P_i \to P_j)$ be the classical probability of a walker moving from point $P_i$ to $P_j$.
        If $P_i$ and $P_j$ are non-adjacent, $p_C(P_i \to P_j)$ represents the sum over the probabilities of all possible paths between those end-points.
        For conciseness, a walker starting at the input node $P_0$ reaches the \textit{Drop} and \textit{Thru} ports with the probabilities $p_C^D$ and $p_C^T$, respectively.
        
        $p_C^D$ and $p_C^T$ can be calculated analytically using two general approaches:
        1) by explicitly summing the probabilities associated with all possible paths from $P_0$ to $P_D$ and $P_T$.
        Note that the number of paths is infinite, as there is no bound on the number of loops the walker can take around each ring.
        As we will see, this is a classical analogue of the quantum-mechanical Feynman path sum calculation;
        2) using a transfer matrix-based Markov chain method.
        As we will see, the first method clarifies the differences between quantum and classical walks over these graphs.
        In contrast, the second method is systematic and practical for systems of multiple rings.

    \subsubsection{Summing over paths}

\begin{figure}
            \includegraphics[width=7.9cm]{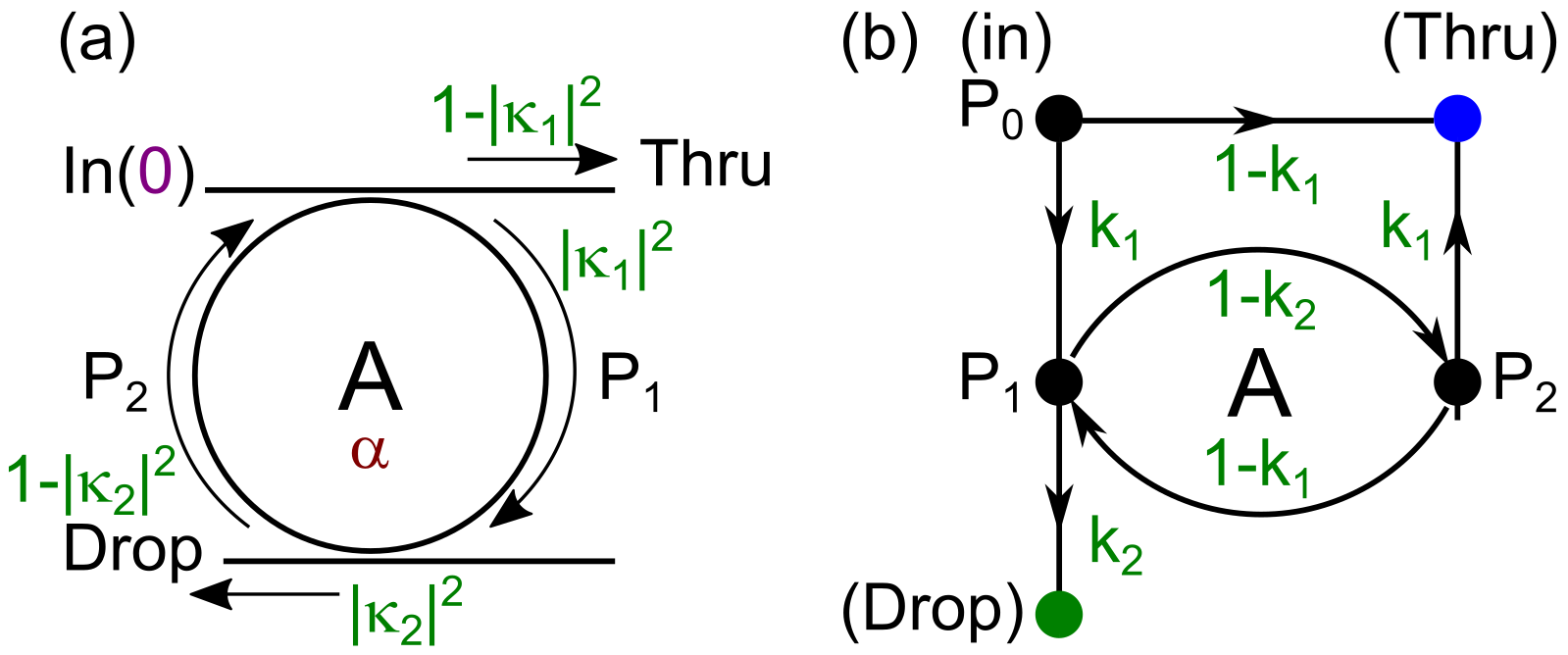}
            \caption{\label{fig:concept:1ring}
            Schematic of the random walk on a single-ring system, and its model.
            (a) Single-ring waveguide configuration, coupling $\kappa_i, i = \{1,2\}$, and loss $\alpha$ coefficients.
            (b) Graph used to model random walks for the single-ring system. $k_i = |\kappa_i|^2, i = \{1,2\}$ are the walker hopping probabilities between the node pairs indicated. The loss coefficient $\alpha$ is omitted in the figure but considered in the calculations.
        }
        \end{figure}

        Here we will consider the sum-over-paths solution to the simplest configuration of a single ring coupled to two linear waveguides, see  Fig.~\ref{fig:concept:1ring}.
        Let us begin by analyzing the probability to reach the \textit{Drop} port.
        Starting at node $P_0$, the possible paths to reach $P_D$ differ only in the number of loops the walker takes around the ring.
        In particular, the walker can reach it after $1/2, 3/2, 5/2, \dots$ turns around the ring, where the associated transmission coefficient is reduced by an $\alpha^\frac{1}{2}$ factor for each half turn around the ring.
        
        To simplify notation, let us consider $t_i = 1-k_i$, and use $n$ to represent the number of random walk steps.
        Also, let $f_C(n)$ describe the classical probability of the walker, having started at node $P_0$, getting to one of the output nodes after $n$ steps.
        Note the difference between $f_C(n)$ and the cumulative probability $p_C(n)$ of the walker having arrived at $P_D$ or $P_T$ in any number of times steps up to $n$.
        Since the walker cannot return from the output nodes $P_D$ or $P_T$ back to any other graph node, the cumulative probabilities $p_C^D(n)$ and $p_C^T(n)$ increase monotonically with $n$, and are determined by summing the probability of the walker having reached the particular output after $n$ steps.

        The walk begins at $n=1$.
        At this stage, the walker either hops directly to the \textit{Thru} port (in which case the walk is over) or enters the ring.
        Hence, $f_C^T(1) = t_1$ and $f_C^D(1) = 0$.
        At $n=2$, provided the walker entered the ring, it can either hop to the \textit{Drop} port or remain in the ring.
        Hence, $f_C^T(2) = 0$ and $f_C^D(2) = k_1k_2\alpha^\frac{1}{2}$.
        Similarly, at $n = 3$, the walker can hop to the \textit{Thru} port but not to the \textit{Drop} port.
        Hence, $f_C^T(3) = t_1 + k_1^2t_2\alpha$ and $f_C^D(3) = 0$.
        As the walk progresses, the probability functions $f_C^D(n)$ and $f_C^T(n)$ can be expressed as
        
        \begin{multline}
            f_C^D(n) = \left\{
            \begin{array}{ll}
                0 & ,n = 1,3,5,\dots \\
                k_1k_2\alpha^\frac{1}{2}(t_1t_2\alpha)^{\frac{n-2}{2}} & , n = 2,4,6,\dots
            \end{array}
            \right.
        \end{multline}
        
        \noindent and
        
        \begin{multline}
            f_C^T(n) = \\ \left\{
            \begin{array}{ll}
                t_1 &, n = 1\\
                0 & ,n = 2,4,6,\dots \\
                t_1+k_1^2t_2\alpha(t_1t_2\alpha)^\frac{n-3}{2} & , n = 3,5,7,\dots
            \end{array}
            \right.
        \end{multline}
        
        The cumulative probability $p_C^D(n)$ of the walker reaching the \textit{Drop} port after $n$ steps can be explicitly calculated as a geometric series sum.
        For $n = 2,4,6,\dots$:
        
        \begin{align}
            p_C^D(n)
            = & 
                \sum 
                _j^n f_C^T(j)
                \\
            = & k_1 k_2\alpha^\frac{1}{2} \left[1+t_1 t_2\alpha
                + \dots
                + (t_1 t_2\alpha)^\frac{n-2}{2}\right]
                \label{eq:CRW:1ring:po-pd-sum}
        \end{align}
        
        The total (cumulative) probability after an infinite number of steps $p_C^D \equiv p_C^D(n\to\infty)$ can thus be written as
        
        \begin{align}
            p_C^D = k_1 k_2\alpha^\frac{1}{2}\sum_{m=0}^{\infty} (t_1 t_2\alpha)^m
            = & \frac{k_1 k_2\alpha^\frac{1}{2}}{1-\alpha t_1 t_2},
            \label{eq:CRW:1ring:p0-pd-final}
        \end{align}
        
        \noindent where $m = (n-2)/2$, for $n = 2,4,6,\dots$.
        
        Similarly, for the \textit{Thru} port, for $n = 1,3,5,\dots$:
        
        \begin{align}
            p_C^T(n) 
            = & 
                \sum 
                _j^n f_C^T(j)
                \\
            = & t_1 + \alpha k_1^2 t_2 \left[ 1+\alpha t_1 t_2
                + \dots
                +(\alpha t_1 t_2)^\frac{n-3}{2} \right]
                \label{eq:CRW:1ring:po-pt-sum}
        \end{align}
        
        The total (cumulative) probability after an infinite number of steps $p_C^T \equiv p_C^T(n\to\infty)$ can thus be written as
        
        \begin{equation}
            p_C^T
            = t_1 + \alpha k_1^2 t_2 \sum_{m=0}^\infty (\alpha t_1 t_2)^m
            = \frac{t_1 + t_2\alpha - 2t_1 t_2 \alpha}{1 - t_1 t_2 \alpha},
            \label{eq:CRW:1ring:p0-pt-final}
        \end{equation}
        
        \noindent where $m = (n-3)/2$ for $n = 3,5,7,\dots$.
        It is easy to check that $p_C^T + p_C^D = 1$ for the lossless case ($\alpha = 1$).
        
        As the number $n$ of walk steps increases, the $f_C(n)$ functions alternate between zero and non-zero probabilities, depending on the parity.
        This leads $p_C^D(n)$ and $p_C^T(n)$ to increase monotonically with $n$, with non-zero increases respectively for even and odd $n$.
        
        This method mirrors the quantum-mechanical calculation we will show later by explicitly adding the probabilities associated with all possible paths.
        Nevertheless, considering all possible paths becomes unwieldy as a calculation method, as shown in  \textit{Appendix}~\ref{ap:CRW:2rings}, where we use it to calculate the probabilities associated with the two-ring configuration.
        A more convenient approach relies on the Markov Chain method, described next.
        
    \subsubsection{Markov Chain method}

        Let the tuple $p^{(n)}$ describe a list of probabilities that the walker will be found in each of the nodes after $n$ steps:
        
        \begin{equation}
            p^{(n)} = T^np^{(1)}
        \end{equation}
        where $p^{(1)}$ describes the initial probability distribution of the walker, and $T^n$ is the transfer matrix describing the node transition probabilities after $n$ steps. Now, note that the same matrix $P$ diagonalizes $T^n$ for any power $n$: 
        \begin{equation}
            T=PDP^{-1} \to T^n=(PDP^{-1})^n=P D^n P^{-1}.
        \end{equation}
        So
        \begin{equation}
            D^n = P^{-1}T^nP.
        \end{equation}
        
        The transfer matrix $T^1$ describes the hopping probabilities corresponding to the first step and can be directly read out from the graph of Fig. \ref{fig:concept:1ring}. We can then obtain the matrix $P$ that diagonalizes $T$, i.e., find $P$ and diagonal $D$ such that $D=P^{-1}TP$, and calculate the probability distribution over the nodes after $n$ steps as
        \begin{equation}
            \label{eq:Markov:pNmain}
            p^{(n)} = T^n p^{(1)}=P D^{n} P^{-1} p^{(1)}
        \end{equation}
        
        We can use Eq.~(\ref{eq:Markov:pNmain}) to find the probability distribution for the walker for any number of rings and any number $n$ of steps, with an arbitrary initial probability distribution of the walker over the nodes.
        For the case of the quantum random walk that we will describe later, the approach will be very similar, except that the tuples represent probability amplitudes rather than probabilities, with $T$ encoding transition amplitudes instead.
        
        For the graph corresponding to the single-ring configuration of Fig.~\ref{fig:concept:1ring}(b) and a walker starting at $P_0$, we can write

        \begin{eqnarray}
            p_C^{(n)} = \begin{pmatrix} p^0\\p^1\\p^2\\p^D\\p^T \end{pmatrix}_C^{(n)}
            = T^n \begin{pmatrix} 1\\0\\0\\0\\0 \end{pmatrix},
            \label{eq:Markcov:classical:P}
            \\
            T^1 =
            \begin{pmatrix}
            0	&	0	            &	0	            &	0	&	0	\\
            k_1	&	0	            &	t_1\alpha^\frac{1}{2}&	0	&	0	\\
            0	&	t_2\alpha^\frac{1}{2}&	0	            &	0	&	0	\\
            0	&	k_2\alpha^\frac{1}{2}&	0	            &	1	&	0	\\
            t_1	&	0	            &	k_1\alpha^\frac{1}{2}&	0	&	1	
            \end{pmatrix},
            \label{eq:Markcov:classical:T}
        \end{eqnarray}
        where the $\alpha^\frac{1}{2}$ terms account for the losses in each half-ring.
        Plugging in Eqs.~(\ref{eq:Markcov:classical:P}) and (\ref{eq:Markcov:classical:T}) in Eq.~(\ref{eq:Markov:pNmain}) for $n\to\infty$ recovers the $p_C^D$ and $p_C^T$ obtained in the previous section, Eqs.~(\ref{eq:CRW:1ring:p0-pd-final}) and (\ref{eq:CRW:1ring:p0-pt-final}), respectively.
        
\begin{figure}
            \centering
            \includegraphics[width=8.6cm]{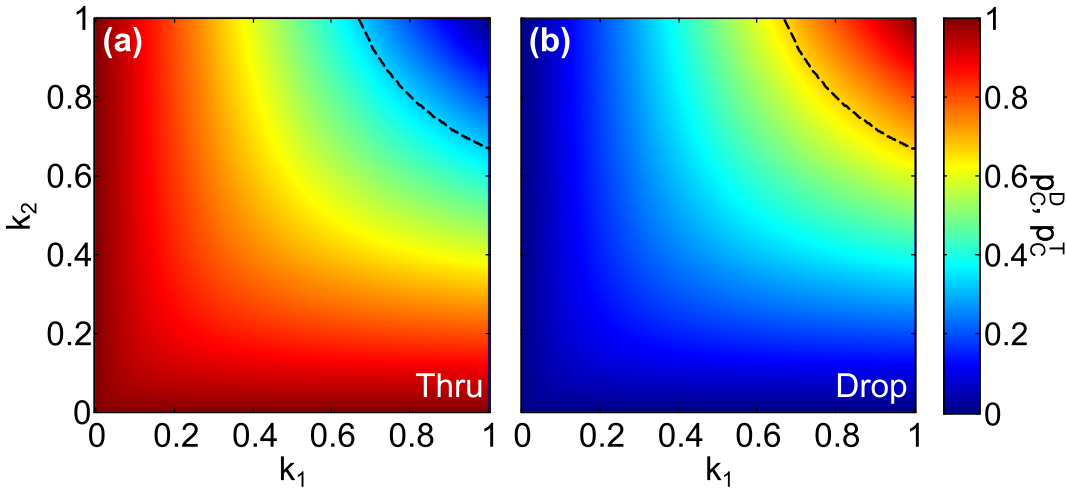}
            \caption{Probabilities of getting to the \textit{Thru} and \textit{Drop} ports for the classical random walk on the single-ring graph of Fig. \ref{fig:concept:1ring}(b).
            (a,b) Probability of reaching $P_T$ (a) and $P_D$ (b), as a function of the two hopping probabilities $k_1$ and $k_2$.
            }
            \label{fig:CRW}
        \end{figure}

        To help compare the different random walk dynamics, let us set $p_g=2/3$ as an arbitrary threshold probability for $P_D$, and analyze the parameter regimes that are sufficient to reach that threshold. Fig.~\ref{fig:CRW} shows the $p_C^T$ and $p_C^D$ distributions for the classical random walk, sweeping over the hopping probabilities $k_1$ and $k_2$.
        The black dashed lines delimit the $k_1,k_2$ region for which the goal is achieved, i.e., $p_C^D > p_g=2/3$.
        The $(k_1,k_2)$ region of success (where $p_C^D>p_g=2/3$) is relatively small, and $p_C^D = 1$ only for $k_1=k_2=1$.
        Since $p_C^T$ and $p_C^D$ add up to one for the lossless case ($\alpha=1$), all \textit{Thru} port observations are complementary to the \textit{Drop} and thus all $p_C^T$ results shall henceforth be omitted from the figures.

    \subsection{Quantum Random Walk}
        \label{sec:QRW}

        While the final probability distributions of a classical walk depend only on the initial state and the hopping probabilities $k_i$, photons display constructive and destructive interference effects resulting in different dynamics.
        For ease of comparison with the classical case, let us assume all ring waveguides are identical, having the same radii and index of refraction. The probability amplitude acquired after each ring round is 
        \begin{equation}
            \label{eq:QRW:gamma}
            \gamma=\alpha e^{i \theta},
        \end{equation}

        The loss coefficient $\alpha = e^{-(\alpha_{t} (L + L_c) + \alpha_{b}L)}$ depends on the material absorption $\alpha_{t}$ and waveguide bending loss $\alpha_{b}$, where $L = 2\pi r$ and $L_c$ are the ring and coupler lengths ($L_c > 0$ for a \textit{racetrack} ring system).
        The phase $\theta$ acquired by a photon after going around the loop once is given by
        \begin{equation}
            \label{eq:QRW:theta}
            \theta = 2\pi n_{eff} \frac{L}{\lambda},
        \end{equation}
        \noindent where $n_{eff}$ and $\lambda$ are the waveguide effective refractive index and photon wavelength, respectively. The assumption that all rings have the same radius will allow us to characterize the \textit{Drop} and \textit{Thru} probabilities as a function of time, parameterized in units corresponding to the time taken for light to go around one half-ring, $\Delta t= \pi r n_{eff}/c$.
        As we will see, interference effects will change the \textit{Drop} and \textit{Thru} probabilities and can be tuned by changing the photon wavelength $\lambda$, or ring radius $r$ (in the latter case, the unit of time will also change accordingly). 
        
        There are several ways to obtain the probability that a quantum walker starting at the input port $P_0$ will arrive at the \textit{Drop} and \textit{Thru} ports,  respectively $p_Q^D$ and $p_Q^T$.
        Again, in this section, we will obtain those for the single-ring configuration, leaving the case of two or more rings for discussion in Appendix~\ref{ap:QRW:2rings}.
        
        The first method is the Feynman path sum over amplitudes corresponding to all possible paths. We refer to the Fig.~\ref{fig:concept:1ring}(b) schematic, with the understanding that now we must represent probability amplitudes $a(P_i \to P_j)$ for the different transitions, rather than hopping probabilities.
        The phases must be chosen so that the couplings between waveguides is described by a unitary transformation;
        for the single-ring configuration, the simplest choice is to set all amplitude phases equal to zero, except $a(P_0 \to P_1)= -k_1^\frac{1}{2}$.
        As in the classical case, we define the probabilities $f_Q^D(n)$ and $f_Q^T(n)$ of the walker getting to either output port at step $n$:
        \begin{multline}
            f_Q^D(n) =\\ \left\{
            \begin{array}{ll}
                0 & ,n = 1,3,5,\dots \\
                -(k_1k_2\gamma)^\frac{1}{2}\left((t_1t_2)^\frac{1}{3}\gamma\right)^{\frac{n-2}{2}} & , n = 2,4,6,\dots
            \end{array}
            \right.
        \end{multline}
        and
        \begin{multline}
            f_Q^T(n) =\\ \left\{
            \begin{array}{ll}
                t_1^\frac{1}{2} &, n = 1\\
                0 & ,n = 2,4,6,\dots \\
                t_1^\frac{1}{2}-k_1t_2^\frac{1}{2}\gamma\left((t_1t_2)^\frac{1}{2}\gamma\right)^\frac{n-3}{2} & , n = 3,5,7,\dots
            \end{array}
            \right.
        \end{multline}
        
        The cumulative probability amplitudes of getting to the output ports in any number of steps can be obtained by mirroring the calculation for the classical case, except now we are summing amplitudes.
        For the \textit{Drop} case:
        \begin{align}
            a^D(n)
            = & 
                \sum 
                _j^n f_Q^D(j)
                \\
            = & -(k_1 k_2 \gamma)^\frac{1}{2} \left[1 + (t_1 t_2)^\frac{1}{2}\gamma
                + \dots + \right.\nonumber\\
                & \hspace{3.625cm}  \left. + ((t_1 t_2)^\frac{1}{2}\gamma)^\frac{n-2}{2} \right]
            \label{eq:QRW:1ring:po-pd-sum_a}
        \end{align}
        
        The total (cumulative) probability amplitude $a^D \equiv a^D(n\to\infty)$ after an infinite number of steps can thus be written as
        
        \begin{equation}
            a^D= - (k_1 t_2\gamma)^\frac{1}{2} \sum_{m=0}^\infty\left((t_1t_2)^\frac{1}{2}\gamma\right)^m
            = -\frac{(k_1 k_2 \gamma)^\frac{1}{2}}{1- (t_1 t_2)^\frac{1}{2}\gamma}. \label{eq:qdrop1ring}
        \end{equation}
        
        \noindent where $m = (n-2)/2$ for $n = 2,4,6,\dots$.
        The calculation for $a^T$ also follows the same line as the classical sum over paths calculation:
        
        \begin{align}
            a^T(n)
            = & 
                \sum
                _j^n f_Q^T(j)
                \\
            = & t_1^\frac{1}{2} - k_1 t_2^\frac{1}{2}\gamma \left[1
                + (t_1t_2)^\frac{1}{2}\gamma
                + \dots + \right. \nonumber\\
                &  \hspace{3.2cm} + \left.\left((t_1t_2)^\frac{1}{2}\gamma\right)^\frac{n-3}{2} \right] 
            \label{eq:QRW:1ring:po-pt-sum_a}
        \end{align}
        
        The cumulative probability amplitude of getting to the \textit{Thru} port then becomes
        
        \begin{align}
            a^T = t_1^\frac{1}{2} - k_1 t_2^\frac{1}{2}\gamma \sum_{m=0}^\infty\left((t_1t_2)^\frac{1}{2}\gamma\right)^m
            = \frac{t_1^\frac{1}{2} -t_2^\frac{1}{2}\gamma}{1 - (t_1 t_2)\frac{1}{2}\gamma},
            \label{eq:qt1ring}
        \end{align}
        
        \noindent where $m = (n-3)/2$ for $n = 3,5,7,\dots$. We see that the \textit{Drop} and \textit{Thru} probabilities for the classical case (and probability amplitudes for the quantum case) are both given as a sum over probabilities (or amplitudes) corresponding to each of the infinite possible paths from $P_0$ to $P_D$ and $P_T$.
        We can obtain the expressions for the quantum amplitudes $a^D$ and $a^T$ by substituting amplitudes for probabilities in the classical walk result, as shown by a comparison between Eqs.~\ref{eq:CRW:1ring:p0-pd-final} and \ref{eq:qdrop1ring}, and Eqs.~\ref{eq:CRW:1ring:p0-pt-final} and~\ref{eq:qt1ring}, respectively.
        The substitutions are:
        \begin{align}
            k_i \to     & k_i^\frac{1}{2}e^{i \phi_i}\\
            t_i \to     & t_i^\frac{1}{2} \label{eq:CRWvsQRW:subs:t_i}\\
            \alpha \to  & \gamma = \alpha e^{i\theta}
        \end{align}
        
        \noindent where the phases $\phi_i$ must be chosen to guarantee unitarity of all waveguide couplings.
        
        To obtain the quantum probabilities of a walker going from $P_0$ to $P_D$ or $P_T$, as always in quantum mechanics, we must take the absolute value squared of the amplitudes obtained above: 
        \begin{align}
             p_Q^D = &
             \frac{k_1 k_2\alpha^\frac{1}{2}}{1 + t_1 t_2 \alpha - 2(t_1 t_2)^\frac{1}{2} \alpha \cos \theta}
             \label{eq:QRW:1ring:Id/I0}
             \\
             p_Q^T = &
             \frac{t_1 + t_2\alpha - 2(t_1 t_2)^\frac{1}{2} \alpha \cos \theta}{1 + t_1 t_2 \alpha - 2(t_1 t_2)^\frac{1}{2} \alpha \cos \theta}
             \label{eq:QRW:1ring:It/I0}
        \end{align}
        For the case of no loss ($\alpha=1$), it is easy to check that  $p_Q^D + p_Q^T= 1$, as expected.
        
        The expressions above give us the quantum probabilities of reaching either output port after an arbitrarily long time.
        Suppose we want to know this probability after $n$ random walk steps.
        In that case, we can truncate the sums above in the corresponding term, or equivalently, use the Markov chain approach described in Appendix~\ref{ap:QRW:1ring:Markov}.
        
        Another way of obtaining the steady-state \textit{Drop} and \textit{Thru} probabilities involves appealing to the classical/quantum correspondence principle.
        The intensity ratios predicted by the classical electrodynamic description of the problem must match the quantum mechanical probability calculations, so $p_Q^D=I_D/I_0$ and $p_Q^T=I_T/I_0$, where $I_0, I_D, I_T$ represent the input, \textit{Drop} and \textit{Thru}
        steady-state
        intensities predicted by classical electrodynamics.
        This way of doing the calculation allows us to use the boundary conditions at each coupling region to solve Maxwell's equations, and from the solution, obtain the quantum-mechanical result.
        The quantum probabilities for \textit{Drop} and \textit{Thru} ports we obtained above match previously reported results \cite{Rabus2007}.

        \begin{figure}
            \includegraphics[width=8.6cm]{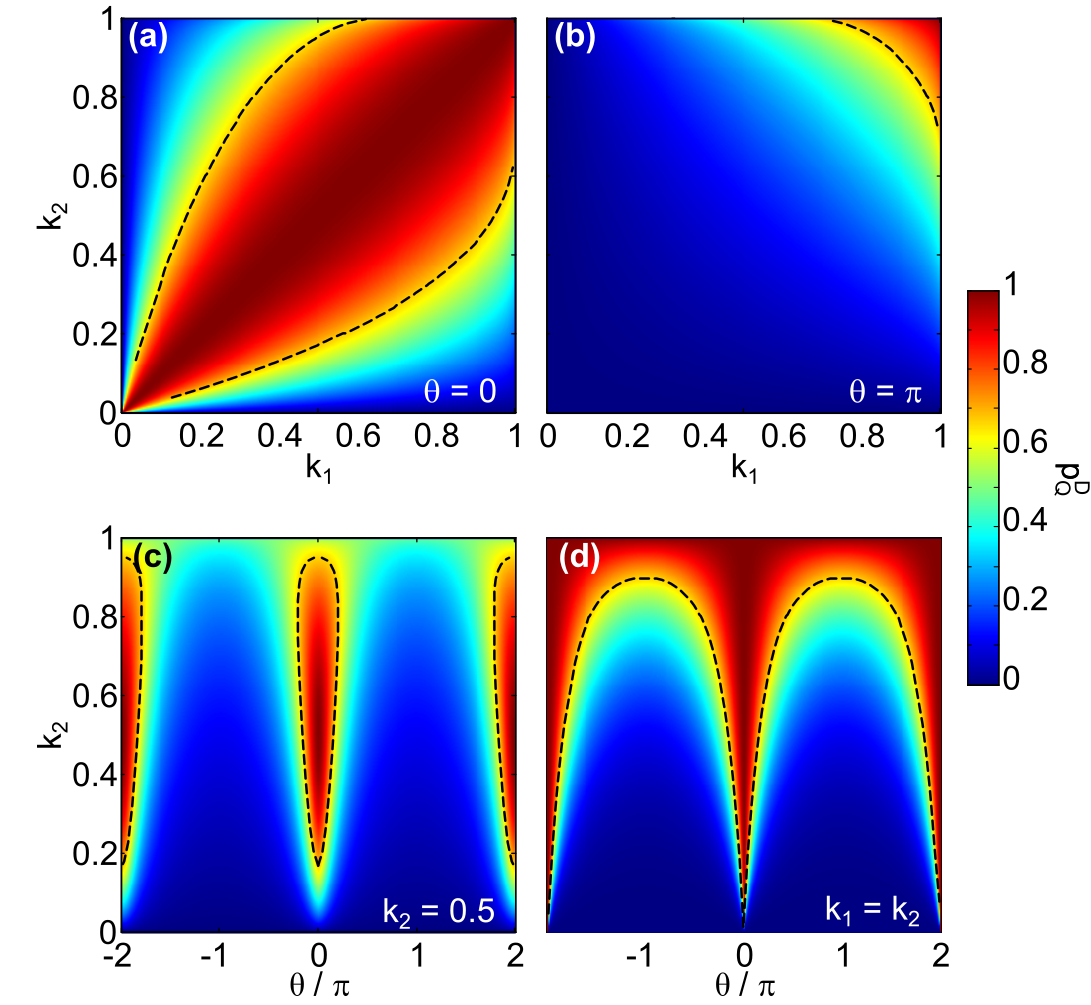}
            \caption{Quantum random walk implemented by coherent light propagation in a single-ring resonator.
            (a-b) Probability $p_Q^D$ of reaching the \textit{Drop} port, sweeping over hopping probabilities $k_1,k_2$ and single-ring acquired phase $\theta = 0$ (a) and $\theta = \pi$ (b).
            (c-d) $p_Q^D$ for sweeping $k_1$ and $\theta$, with fixed $k_2 = \frac{1}{2}$ (c) and for matching $k_2=k_1$ (d).
            The black dashed lines indicate the goal-hitting threshold for $p_Q^D = 2/3$.
            All plots share the same intensity color scale.
            }
            \label{fig:QRW:PD}
        \end{figure}
        
        Fig.~\ref{fig:QRW:PD} plots $p_Q^D$ for different combinations of $k_1,k_2$ and $\theta$.
        Unlike the classical random walk, the response depends strongly on the new parameter $\theta$; the phase acquired after one loop around the ring. The goal-hitting region in parameter space, corresponding to $p_Q^D > p_g=2/3$, is shown in Fig.~\ref{fig:QRW:PD}(a,b) for $\theta = \{0,\pi\}$, respectively.
        As expected, the resonant ($\theta = 2\pi n, n \in \mathbb{Z}$) and antiresonant ($\theta = \pi + 2\pi n$) phases give maximum and minimum conditions for  $p_Q^D$.
        The $\theta$ sweeps from Fig.~\ref{fig:QRW:PD}(c-d) exhibit the coherent phase effect by sweeping over $k_1$, with a fixed $k_2 = 1/2$ value (Fig.~\ref{fig:QRW:PD}(c)) and the matching $k_1=k_2$ condition (Fig.~\ref{fig:QRW:PD}(d)).
        Again the latter is confirmed to maximize $p_Q^D$, and thus the goal-hitting chance.
        
        The observations from Figs.~\ref{fig:CRW} and \ref{fig:QRW:PD} indicate that the coherent phase $\theta$ can be tuned to increase the goal-hitting chance. In the next section, we explore the comparison between quantum and classical random walks on coupled ring resonators in more detail.

    \subsection{Comparison between classical and quantum random walks}
        \label{sec:CRWvsQRW}
        In the previous two sections, we calculated the classical and quantum solutions for the random walk on the graph of Fig.~\ref{fig:concept:1ring}(b), which models a ring resonator.
        Quantum random walks can achieve faster propagation than classical walks over different graphs.
        In this section, we consider the goal of traversing the graph from the starting node $P_0$ to the \textit{Drop} port $P_D$, and compare the performance of quantum and classical walks, for any finite number of steps $n$, but also in the steady-state, obtained as the number of steps $n \to \infty$.

    \subsubsection{Steady-state solutions}

        Section~\ref{sec:CRW} demonstrates that apart from the loss coefficient $\alpha$, the only tunable parameters for the classical random walk are the hopping probabilities $k_i$, c.f. Eqs.~\ref{eq:CRW:1ring:p0-pd-final} and \ref{eq:CRW:1ring:p0-pt-final}.
        On the other hand, the quantum random walk dynamics over the same structure depend not only on the coupling coefficients but also on the ratio between ring radius $r$ and wavelength $\lambda$, which can be conveniently parameterized by the phase $\theta$ acquired after a single ring round-trip.
        Even the loss coefficient $\alpha$, which in the classical regime only harms the goal-hitting chance, \textit{can} sometimes be used to optimize the ring resonance contrast \cite{Rabus2007} in the quantum regime.
        
\begin{figure}
            \centering
            \includegraphics[width=8.6cm]{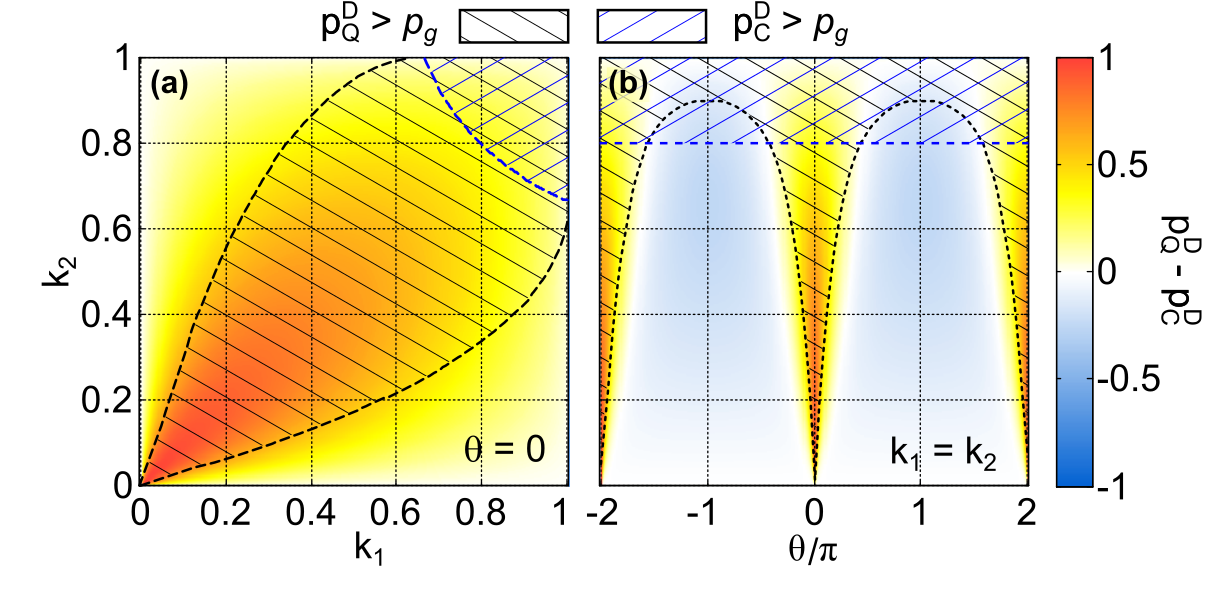}
            \caption{Comparison between classical and quantum probabilities of reaching the \textit{Drop} port of the single-ring resonator. 
            (a-b) Difference between quantum $p_Q^D$ and classical $p_C^D$ distributions, sweeping over coupling coefficients $k_1,k_2$ with a fixed coherent phase $\theta = 0$ (a), and sweeping $k_1, \theta$, for matching $k_2=k_1$ (b).
            The black and blue patterns highlight the parameter combinations for which $p_Q^D$ and $p_C^D$ overcome the goal-hitting threshold $p_g=2/3$.
            The color scale is shared by both images.
            }
            \label{fig:QRW-CRQ:comparison}
        \end{figure}

        The steady-state comparison between the classical and quantum regimes can be done by analyzing the results from Eqs.~\ref{eq:CRW:1ring:p0-pd-final} and \ref{eq:QRW:1ring:Id/I0}, once more focusing on the lossless case $\alpha = 1$.
        Comparison between $p_C^D$ and $p_Q^D$ immediately shows that for the resonant condition $\theta = 2\pi n, n\in \mathrm{Z}$, $p_Q^D \geqslant p_C^D$ for any choice of $(k_1,k_2)$.
        Fig.~\ref{fig:QRW-CRQ:comparison} shows the difference between the quantum and classical probabilities of reaching the \textit{Drop} port $p_Q^D-p_C^D$ as a function of the coupling coefficients (Fig.~\ref{fig:QRW-CRQ:comparison}(a)) and as a function of the round-ring acquired phase $\theta$ for matching coupling constants (Fig.~\ref{fig:QRW-CRQ:comparison}(b)).
        The color-scale shows parameter combinations for quantum-over-classical advantage outperforms in shades of orange and vice versa in shades of blue.
        The black and blue line patterns highlight the parameter combinations for which the probability of reaching the \textit{Drop} ports is larger than the chosen goal-hitting threshold $p^D > p_g=2/3$.
        
        Since Fig.~\ref{fig:QRW-CRQ:comparison}(a) is obtained for the resonant condition $\theta = 0$, $p_Q^D \geqslant p_C^D$ for all $(k_1, k_2)$ combinations.
        On the other hand, Fig.~\ref{fig:QRW-CRQ:comparison}(b) shows that for non-resonant conditions, $p_Q^D$ can often be lower than the classical counterpart $p_C^D$, with a difference minimum of $\approx -0.25$.
        Further, for $(k_1 = k_2) > 4/5$ and $\theta \in [\pi/2 + m,3\pi/2+m],\ m\in\mathbb{Z}$, the classical walk outperforms its quantum counterpart \textit{and} overcomes the goal-hitting threshold ($p_C^D > p_Q^D$ and $p_C^D > p_g=2/3$).

        This oscillation of $p_Q^D$ is a wave phenomenon, resulting in a quantum \textit{Drop} probability that may be either higher or lower than the classical counterpart. A simple calculation shows that in the absence of losses ($\alpha=1$), if we average  $p_Q^D$ over the acquired phase $\theta$, we recover the classical result $p_C^D$ (lossless case):  
        
        \begin{align}
           \left< p_Q^D \right>
           = & \frac{1}{2\pi}\int_0^{2\pi} p_Q^D(\theta) \mathrm{d}\theta\\
           = & \frac{1}{2\pi}\int_0^{2\pi} \frac{k_1k_2}{1+t_1t_2-2(t_1t_2)^\frac{1}{2}\cos \theta} \mathrm{d}\theta\\
           = & \frac{k_1k_2}{1-t_1t_2}
           = p_C^D, \label{eq:QRW-average-Id}
        \end{align}
        As $p_Q^D+p_Q^T=1$, averaging $p_Q^T$ over $\theta$ will also recover the classical \textit{Thru} probability.

        This simple calculation shows how the classical behavior is recovered from the quantum behavior as an average over the only quantum parameter in this model, the phase $\theta$. It is an alternative way of finding the result that is expected if we perform the experiment with incoherent light, for which there will be no definite phase acquired after each ring round trip.

    \subsubsection{Time-domain solutions and goal-hitting time}

        \begin{figure}
            \centering
            \includegraphics[width=8.6cm]{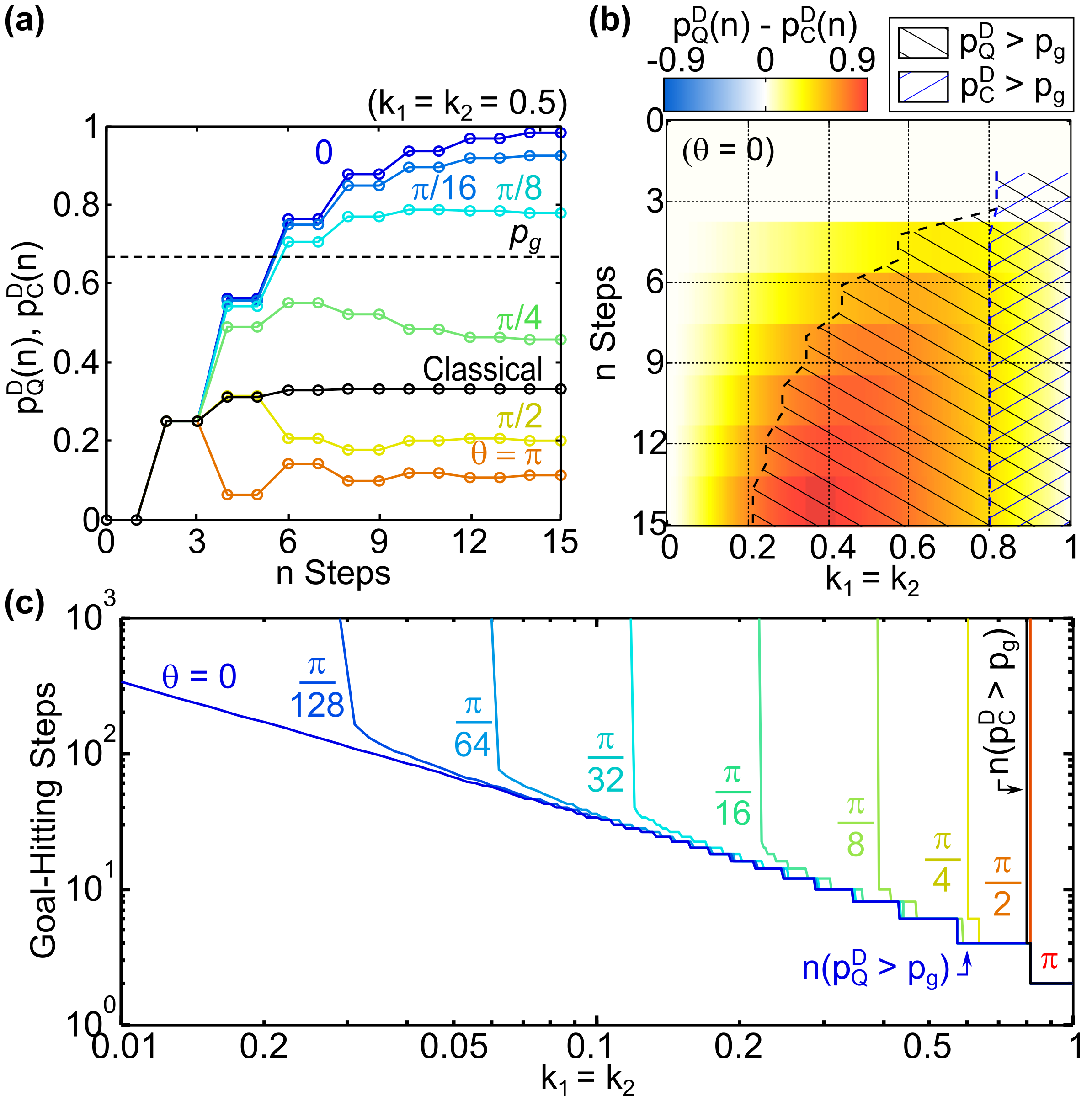}
            \caption{Comparison between the dynamic classical ($p_C^D$) and quantum ($p_Q^D$) random walk probabilities of reaching the \textit{Drop} port of the single-ring resonator.
            (a) Dynamic $p_C^D$ (black) and $p_Q^D$ (color) over the random walk step $n$ for variable coherent phases $\theta$ and fixed coupling coefficients $k_1=k_2=1/2$.
            Horizontal dashed line: Goal-hitting threshold $p_g$.
            (b) Difference between $p_Q^D$ and $p_C^D$ distributions over the random walk step, as a function of
            the matched coupling coefficients $k_1 = k_2$.
            (c) Log-log plot of the goal-hitting time (defined at $p_g = 2/3$) as a function of $k_1 = k_2$ for the classical (black) and quantum (color) regimes for variable coherent phases $\theta$.}
            \label{fig:QRW-CRW:TimeDomain}
        \end{figure}
        
        Thus far, we have only used Eq.~(\ref{eq:Markov:pNmain}) to analyze the steady-state random walk solutions ($n \to \infty$), whose output probabilities agree with the electromagnetic steady-state solutions for the intensity obtained from Maxwell's equations.
        As long as all rings have the same radius, it is also possible to study the time-dependent solution, corresponding to the output probabilities in multiples of the time it takes light to traverse one half-ring.
        One can read the time-dependent solutions from our formulations for finite $n$: Eqs.~\ref{eq:CRW:1ring:po-pd-sum} and \ref{eq:CRW:1ring:po-pt-sum} for the classical walk and the absolute square of Eqs.~\ref{eq:QRW:1ring:po-pd-sum_a} and \ref{eq:QRW:1ring:po-pt-sum_a} for the quantum.

        Fig.~\ref{fig:QRW-CRW:TimeDomain}(a) shows the time evolution of the \textit{Drop} probabilities for the classical and quantum random walks on a single-ring configuration, considering the balanced $k_1 = k_2 = 1/2$ case, for several values of the coherent phase $\theta$.
        Whereas the classical \textit{Drop} probability converges monotonically to the steady-state solution, the quantum \textit{Drop} probability shows fluctuations that depend on $\theta$.
        The fluctuations become more significant when we are close to the antiresonant condition $\theta \to \pi + 2n\pi, n \in \mathbf{Z}$, and virtually vanish close to the resonant condition $\theta \to 2n\pi, n \in \mathbf{Z}$.
        It is worth recalling that for both classical and quantum walks, the \textit{Thru} and \textit{Drop} probabilities only vary at odd and even time-step values, respectively. 
        For small values of $\theta$ we are close to the resonance condition, and the quantum random walk has a better performance than the classical counterpart. At resonance, and with $k_1=k_2=0.5$, the threshold \textit{Drop} probability goal $p_g=2/3$ is reached in $n=6$ time steps, whereas the classical random walk never reaches this goal, unless $(k_1=k_2) \geqslant 4/5$ (as can be checked with Eq.~\ref{eq:CRW:1ring:po-pd-sum}).

        Fig.~\ref{fig:S:TimeDomain} of \textit{Appendix}~\ref{ap:QRW:1ring} shows the individual classical and quantum \textit{Drop} time evolution for variable $\theta$ and $k_1 = k_2$ conditions. The sum of \textit{Drop} and \textit{Thru} probabilities for finite $n$ does not add up to 1, even in the lossless case, as walkers are still filling the waveguide nodes towards the steady-state regime, reached as $n \to \infty$.

        Fig.~\ref{fig:QRW-CRW:TimeDomain}(b) compares the time evolution between the classical random walk and the resonant quantum random walk for sweeping values of $k_1 = k_2$.
        The black and blue line patterns highlight the parameter space regions for which 
        $p_Q^D(n) \geqslant p_g$ and $p_C^D(n) \geqslant p_g$, respectively, where $p_g = 2/3$.
        One can see that the quantum \textit{Drop} probability is much higher than its classical counterpart around the resonances (as indicated by the orange shade), reaching the goal-hitting threshold $p_g$ after 6 steps for $\theta = 0$, while the classical random walk (for $k_1 = k_2 = 1/2$) never does.
        Also, the quantum walk hits the goal for all $(k_1 = k_2) > 0$ values, with a generally higher \textit{Drop} probability (thus larger goal-hitting chance) than the classical analogue, particularly for low $(k_1 = k_2)$ values.
        Furthermore, the numerical results indicate that in resonance ($\theta = 0$), $p_Q^D(n) \geqslant p_C^D(n)$ for any $k_1=k_2$, meaning that in these conditions, the cumulative probability of the walker having reached the \textit{Drop} port in the resonant quantum random walk regime is never smaller than in the classical case.

        We define the goal-hitting time as the number of walking steps required to achieve the goal-hitting probability threshold $p_g=2/3$.
        Fig.~\ref{fig:QRW-CRW:TimeDomain}(c) plots  the classical and quantum goal-hitting times for variable coherent phase $\theta$ values and $k_1 = k_2 = 1/2$.
        The goal-hitting $k_1 = k_2$ range increases as $\theta$ approximates the resonant condition, taking an increasingly longer hitting time towards the lower $k_1 = k_2$ values.
        Even though Eq.~(\ref{eq:QRW:1ring:Id/I0}) shows that $p_Q^D(\theta = 0) > p_g$ for all $k_1 = k_2>0$, Fig.~\ref{fig:QRW-CRW:TimeDomain}(c) demonstrates that there is an approximate power law between the goal-hitting time and $k_1 = k_2$.

\section{Feasibility Study}
    \label{sec:Experimental}
    
    The control of coherent and resonant effects in \acrshort{rr}s relies strongly on the waveguide materials and fabrication accuracy.
    \acrshort{rr}s have been around for a long time and have become one of the most commonly used photonic integrated chip (PIC) building blocks in Si photonics.
    However, Si is not transparent in the visible wavelength range and thus is usually operated in the infrared around $\lambda = 1.550$~\micron{}.
    Photonic-aimed polymers provide good transparency in the visible and near-infrared wavelength ranges.
    Still, their lower refractive indices (usually between $1.3$ and $1.7$ in the visible \cite{Liu2009}, up to 1.936 \cite{Ritchie2021}) compared to Si ($n_g=3.6\ @\lambda=1.550$~\micron{}) provide weaker mode confinement and thus larger waveguide bending losses.
    As a result, the \acrshort{rr} dimensions must be much larger, which narrows the free spectral range between resonance peaks, thus reducing the control over the device output.
    
    While infrared Si photonic elements are already used routinely in \acrshort{pic}s, polymeric ones are not.
    This, together with the theoretical models we have described, motivated us to perform preliminary feasibility tests of the proposed quantum random walk implementation using polymer materials (EpoCore/EpoClad) in the visible wavelength range.
    In particular, we assess the experimental control on the coupling coefficients $k_i$ by theoretically calculating and experimentally verifying them as a function of the coupler length $L_s$ and distance $d$.
    One can find sample preparation and characterization details in \textit{Appendix}~\ref{ap:QRW:Feasibility}.
    Fig.~\ref{fig:S:exp-Couplers}(a) shows the top-view light scattering of fiber side-coupled polymeric directional couplers ($L_s = 100$~\micron{}) for three coupler distances $d = \{0, 0.14, 2.5\}$~\micron{}.
    The $d$ dependence can be qualitatively observed by the total, partial, and no coupling between the input (top) to the transport (bottom) waveguides.
    
    The coupling coefficients (intensity splitting) are quantitatively characterized by the intensity at the waveguide outputs (measured at the chip edge).
    Fig.~\ref{fig:S:exp-Couplers}(b,c) shows the intensity splitting obtained for coupler distances $d=0.12$ (b) and $d=0.14$~\micron{} (c), as a function of the coupler length.
    The coupling coefficients were theoretically predicted using a mode solver software \cite{Fallahkhair2008} and Eqs.~\ref{eq:ap-exp-Coupler-Lb}-\ref{eq:ap-exp-Coupler-k} of \textit{Appendix}~\ref{ap:QRW:Feasibility}.
    We observe a general agreement with the theory but with large standard deviations that indicate the experimental challenge.

    \begin{figure}
        \centering
        \includegraphics[width=8.6cm]{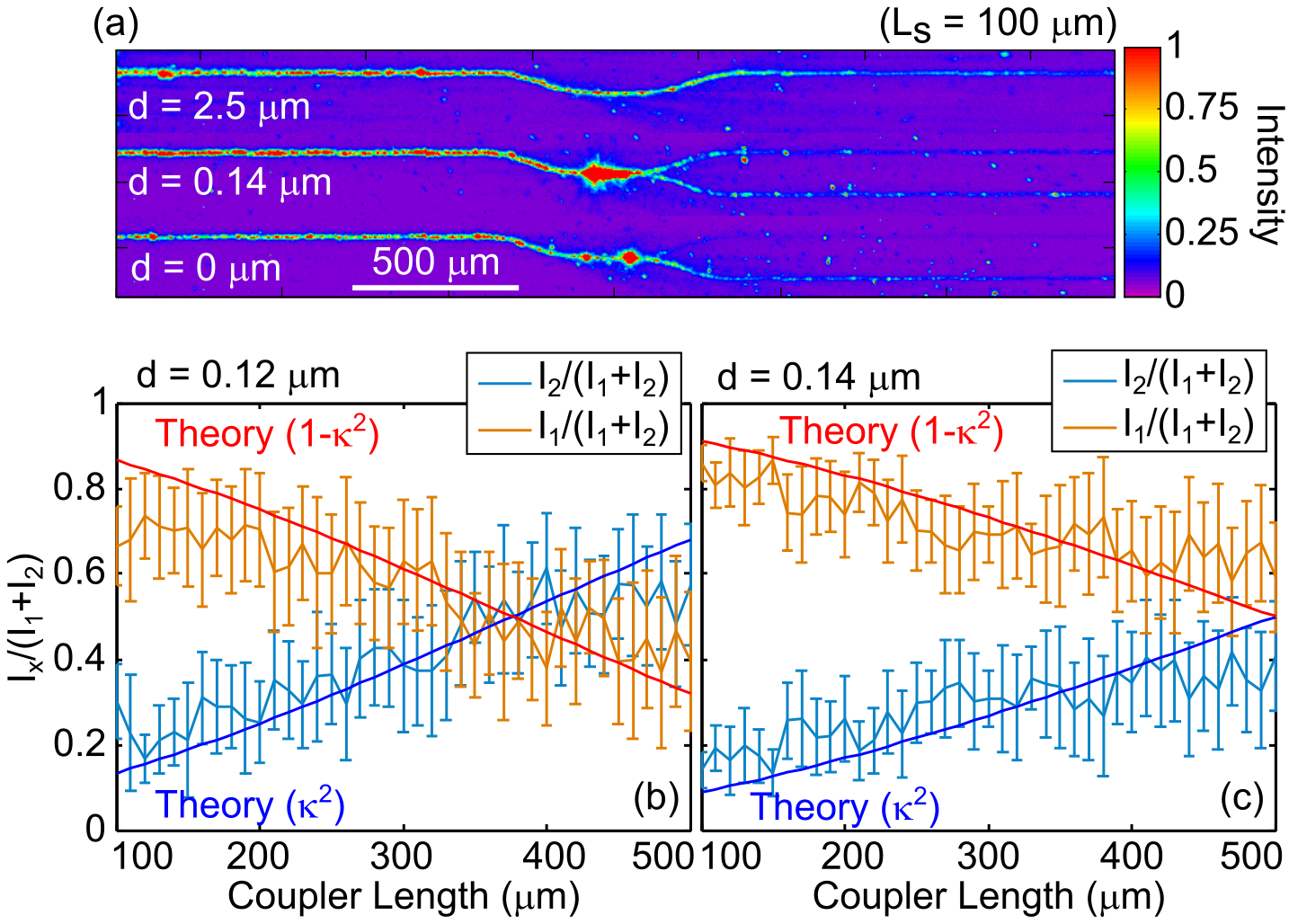}
        \caption{Experimental characterization of the coupling coefficient in parallel planar polymeric waveguides.
        (a) Top-view light-scattering intensity of optical fiber-coupled polymeric-waveguide directional couplers with coupler distances $d = {0 0.14 2.5}$~\micron{}.
        (b,c) Experimental and theoretical intensity splitting (coupling coefficients) as a function of the coupler length for coupler spacing $d=0.12$~\micron{} (a) and $d=0.14$~\micron{} (b).
        The error bars indicate the experimental $\pm$ standard deviation.
        Excitation wavelength $\lambda=635$~nm.
        }
        \label{fig:S:exp-Couplers}
    \end{figure}
    
\begin{figure}
        \centering
        \includegraphics[width=8.6cm]{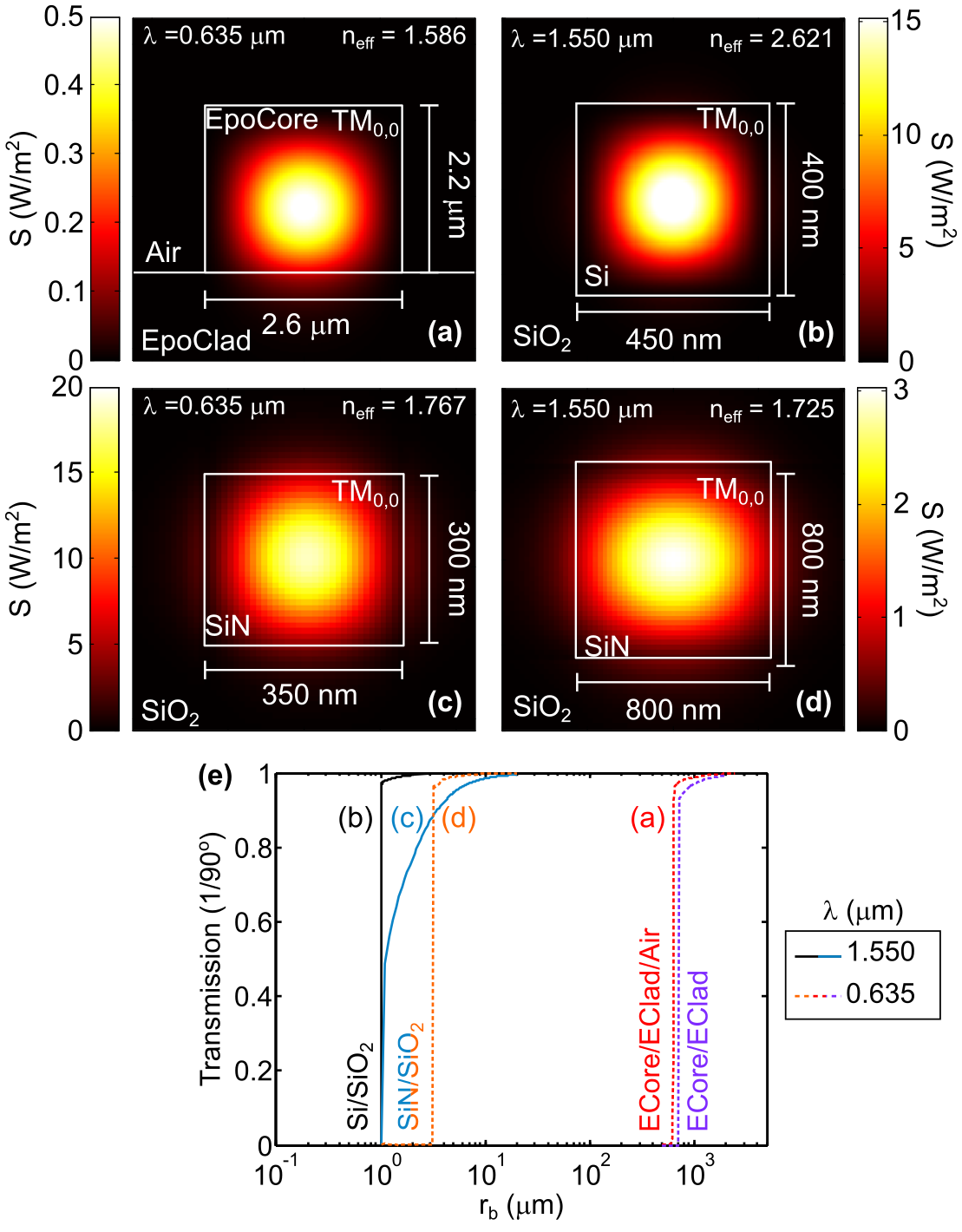}
        \caption{Single-mode waveguiding for different materials and wavelengths.
        (a-d) TM waveguiding mode supported by core/cladding waveguides of EpoCore/EpoClad (a), Si/SiO\tsb{2} (b), and Si\tsb{3}N\tsb{4}/SiO\tsb{2} (abbreviated to SiN/SiO\tsb{2}) (c,d), supporting wavelengths $\lambda = 0.635$~\micron{} (a,c) and $\lambda = 1.55$~\micron{} (b,d).
        (e) Waveguide bending losses (transmission per $90^{\circ}$) associated with (a-d).
        ECore/Eclad (violet curve) relates to an EpoCore core completely surrounded by EpoClad.
        }
        \label{fig:exp:matBeningLosses}
    \end{figure}

    As mentioned above, the maximum free spectral range achievable with a \acrshort{rr} depends on the minimum bending radius supported by the waveguide.
    Fig.~\ref{fig:exp:matBeningLosses} analyzes the simulated bending losses associated with example waveguide configurations with core/cladding materials: EpoCore/Epoclad Fig.~\ref{fig:exp:matBeningLosses}(a), Si/SiO\tsb{2} Fig.~\ref{fig:exp:matBeningLosses}(b) and Si\tsb{3}N\tsb{4}/SiO\tsb{2} (abbreviated to SiN/SiO\tsb{2}) Fig.~\ref{fig:exp:matBeningLosses}(c-d).
    The waveguide designs consist of either a core material on a cladding substrate, surrounded by air (Fig.~\ref{fig:exp:matBeningLosses}(a)), or completely surrounded by the cladding material (Fig.~\ref{fig:exp:matBeningLosses}(b-d)).
    The simulations were performed for wavelengths $\lambda = 0.635$ (Fig.~\ref{fig:exp:matBeningLosses}(a,c)), and $\lambda = 1.55$~\micron{} (Fig.~\ref{fig:exp:matBeningLosses}(b,d)) and the geometries were optimized for single-mode waveguiding in each material-wavelength combination.
    Fig.~\ref{fig:exp:matBeningLosses}(e) plots the bending losses associated with each waveguide design, including the fully surrounded EpoCore/EpoClad configuration.
    The plot shows how the different materials support strikingly different minimum waveguiding bending radii, spanning from $\approx 1$~\micron{} for Si/SiO\tsb{2} up to $700$~\micron{} for EpoCore/EpoClad.
    
    We characterize the bending losses in an EpoCore/ EpoClad/Air configuration and compare them against the simulation (Fig.~\ref{fig:S:exp-BendLosses} of \textit{Appendix}~\ref{ap:QRW:Feasibility}).
    The substantial transmission variance indicates challenging experimental reproducibility beyond the simulation predictions.

\section{Discussion}
    \label{sec:discussion}

    This paper proposes using coupled ring resonators to implement quantum random walks. We model light propagation in series-coupled ring resonators as random walks on an appropriate family of graphs and calculate the finite-time and steady-state solutions for both classical and quantum random walks on these graphs. The probability of a walker reaching either of the two output ports (labeled as \textit{Thru} and \textit{Drop}) can be obtained using a classical random walk formalism by considering the hopping probabilities at each node.
    On the other hand, the physical implementation of the system using \acrshort{rr}s depends on coherent and resonant effects that bias the walker "decisions," leading to modified output probabilities, which are proportional to the intensities expected to be measured experimentally.
    
    The analysis presented has been devoted mainly to the simplest single add-drop photonic ring resonator (\acrshort{rr}) case, which allows drawing the most important observations from the system with a reduced calculation complexity. We have also obtained steady-state solutions for the two-ring configurations.
    Multi-ring systems add complexity and output tunability (as more parameters come into play). Still, the main observation remains that one can use the coherent and resonant \acrshort{rr} effects to tune (quantum bias) the response of a random walk graph and improve its algorithmic efficiency.
    
    The two proposed calculation methods provide an intuitive and a scalable approach to calculate the \textit{Thru} and \textit{Drop} output probabilities.
    While the Feynman path sum becomes unwieldy to tackle systematically, complicated for multi-ring systems, as inter-ring loop combinations need to be considered (see double-ring system analysis in \textit{Appendix}~\ref{ap:RW:2rings}), the Markov chain method provides a platform for straightforward numerical calculation method for any number of rings, providing both steady-state and dynamic probability distributions at each random walk node.
    Analytical solutions can be obtained from the Markov Chain methods using symbolic computational solvers such as \textit{Wolfram Mathematica}.
    \acrfull{fdtd} and \acrfull{fdfd} photonic simulation models can be used to simulate the quantum random walk scenario.
    However, such numerical simulation approaches will require considerable (and perhaps forbidding) computation power and time resources.

    Eq.~\ref{eq:QRW-average-Id} reveals that the coherent phase average of quantum \textit{Drop} and \textit{Thru} port probabilities converge towards the classical solutions. This means the quantum regime corresponds to a field redistribution over the graph, as a function of the coherent effects and ring resonances.
    In terms of goal-hitting rate and speed, there must always be a trade-off between resonant and antiresonant conditions, so that on average, the quantum random walk behaves just like the classical random walk.
    The key is thus to optimize the coherent parameters (ring radii and wavelength) to improve the algorithmic efficiency relative to the classical regime.
    
    It should be noted that even though we have only addressed series-coupled \acrshort{rr} arrays, the same type of modeling can be applied to more complex 2D coupling configurations \cite{Bachman2015}. This suggests that 2D arrays of series- and parallel-coupled ring resonators may provide a flexible, and experimentally accessible model for coherent light propagation in two-dimensional structures.

    
    The control of coherent and resonant effects in \acrshort{rr} relies strongly on the waveguide materials and fabrication accuracy.
    In particular, the effective waveguide refractive index determines the waveguiding mode confinement, which affects single-mode waveguide dimension requirements and its associated bending losses.
    In turn, the bending losses limit the minimum ring radius supported by the waveguide, which plays a key role in the free spectral range between resonance peaks.
    Polymer-based photonic elements have recently arisen as a low-cost, versatile alternative to Si-based photonic integrated chips, with outstanding transparency in the visible and near-infrared wavelength ranges.
    Polymer photonics boast new interesting properties such as biocompatibility, mechanical flexibility, and photosensitive properties that allow \acrfull{dlw}-based 3D patterning.
    However, the lower refractive indices in polymers (usually between $1.4$ and $1.7$ in the visible) compared to Si ($n_g=3.6\ @\lambda=1.550$~\micron{}) provide weaker mode confinement and thus larger waveguide bending losses.
    As a result, the \acrshort{rr} dimensions must be much larger, which narrows the free spectral range, thus reducing the control over the device output.
    Very narrow free spectral ranges (e.g., narrower than the bandwidth of the light source) make it difficult to distinguish resonances experimentally and thus introduce decoherence.
    As the device output becomes the average over multiple wavelength contributions, the fine control described in Section~\ref{sec:QRW} is harmed or lost.
    
    Previous works have reported on the impact of decoherence in quantum random walk systems and the importance of controlling it\cite{Svozilik2012,Biggerstaff2016,Broome2010}.
    Eq.~(\ref{eq:QRW-average-Id}) shows that introducing decoherence (phase average) in the proposed system deviates the output distributions away from the quantum regime and towards the classical random walk solution.
    The effect can be seen as a further output optimization parameter, but also as a constraint, since photonic integrated chip nanofabrication and device operation incur experimental errors, to which narrow free spectral range systems are more sensitive.
    
    Even in perfect experimental conditions, the large minimum radius supported by the polymeric waveguides leads to an extremely reduced free spectral range ($10$-$100$~pm).
    Consequently, accurate experimental characterization would require either an extremely narrow-band light source or a high-precision optical spectrum analyzer.
    The decoherence effects (as a consequence of many averaged resonance steps) would likely frame the polymeric waveguide implementation closer to the classical regime than the quantum.
    Therefore, any of the semiconductor material-based configurations here discussed is more likely to provide successful device implementation than the polymeric ones.

\section{Conclusion} \label{sec:conclusion}
    
    We have proposed series-coupled photonic ring resonators as a platform to implement controllable quantum random walks. We have modeled light propagation in these structures using a suitable family of graphs and analyzed quantum and classical random walk behavior using this model. Light coherence and resonance effects in the quantum random walk provide output tunability based on the wavelength and ring radii.
    Both classical and quantum random walk regimes can be analyzed using Feynman path sums or Markov Chain approaches, where the latter provides an efficient method for both analytical and numerical output probability calculations for graphs with any number of rings.
    The quantum random walk coherent phase average tends towards that of the corresponding classical random walk, imposing a trade-off between resonant and non-resonant conditions.
    When trying to maximize the transport efficiency across the graph, from the initial input node to the \textit{Drop} output node, resonant coherent conditions in a single-ring system allow enhanced transport far beyond its classical counterpart, for a much wider range of coupling constant combinations.
    The time-domain analysis shows slightly slower convergence rates in the quantum domain, but successful goal-hitting for all matching $k_1 = k_2$ conditions, something the corresponding classical walks are incapable of doing for small $k_1=k_2$.
    Mode confinement and waveguide bending loss simulations revealed the importance of material choice and the impact on decoherence effects.
    Preliminary feasibility tests in the visible wavelength range using polymers revealed implementation challenges from the points of view of both the free spectral range between observable resonances and control over device parameters.
    
    The proposed analysis of coupled ring resonators as a platform for quantum random walks has the potential for photonic algorithm implementations that are not restricted to linear ring configurations but could be expanded to complex 2D arrangements. These could be used to experimentally model coherent energy transfer in 2D arrays.

    \begin{acknowledgments}
    We wish to acknowledge Dr. Jérôme Borme for his support on e-beam lithography for the sample nanofabrication.
    We acknowledge funding from the INL Seed Grant project ``Coherent light propagation in photonic quantum walk chips". EFG acknowledges funding of the Portuguese institution FCT – Fundação para Ciência e Tecnologia via project CEECINST/00062/2018. This work was supported by the ERC Advanced Grant QU-BOSS (GA no.: 884676).
    RA acknowledges the Laser Photonics \& Vision Ph.D. program, U. Vigo.
    MCG acknowledges funding of H2020 Marie Skłodowska-Curie Actions (713640)
    \end{acknowledgments}

%

\clearpage
\appendix
\onecolumngrid

\section{Quantum Random Walk on a Single Photonic Ring Resonator}
\label{ap:QRW:1ring}

This appendix gives more details on the quantum random walk on the single-ring configuration modelled the graph of Fig. \ref{fig:concept:1ring}-b. In Section~\ref{ap:QRW:1ring:Markov} we describe the calculation of the \textit{Drop} probability $p_Q^D$ and \textit{Thru} $p_Q^D$ using a the Markov chain approach, which is an alternative to the explicit Feynman path sum discussed in the main text. In Section~\ref{sec:timedomain} we provide more details of the time-domain solution that can be easily obtained using the Markov chain approach.

\subsection{Markov Chain approach}
\label{ap:QRW:1ring:Markov}

Similarly to the classical random walk on the same graph, one can find the eigenmatrix $P$ of the electric field transfer matrix $T$ after $n$ steps (measured in half-ring light paths) so that

\begin{equation}
    D^n = P^{-1}T^nP,
\end{equation}

\noindent where $D^n=\left( D^1 \right)^{\circ n}$ (element-wise power) is the diagonal after $n$ steps, and calculate the electric field amplitudes after any number of steps

\begin{equation}
    \label{eq:Markov:ENmain}
    a^{(n)} = \left(P D^{n} P^{-1}\right) a^{(1)}
\end{equation}

For the quantum random walk on a single ring graph, $T^1$ is given by

\begin{equation}
    T^1 =
    \begin{pmatrix}
    0	                        &	0               	    &	0	                    &	0	&	0	\\
    -(k_1\gamma)^\frac{1}{2}    &	0	                    &	(t_1\gamma)^\frac{1}{2} &	0	&	0	\\
    0	                        &	(t_2\gamma)^\frac{1}{2} &	0	                    &	0	&	0	\\
    0	                        &	(k_2\gamma)^\frac{1}{2}	&	0	                    &	1	&	0	\\
    (t_1\gamma)^\frac{1}{2}     &	0	                    &	(k_1\gamma)^\frac{1}{2}	&	0	&	1	
    \end{pmatrix}
\end{equation}

Finally, the node probability amplitudes after infinite steps can be obtained using Eq.~(\ref{eq:Markov:pNmain}) as

\begin{equation}
    a^{(\infty)} = \begin{pmatrix} a^0\\a^1\\a^2\\a^D\\a^T \end{pmatrix}^{(\infty)}
    = \left(P \left(P^{-1} T^1 P \right)^{\infty} P^{-1}\right)
    \begin{pmatrix} 1\\0\\0\\0\\0 \end{pmatrix}
    = 
    \begin{pmatrix}
    0\\0\\0\\
    -\frac{(k_1 k_2 \gamma)^\frac{1}{2}}{1-(t_1t_2)^\frac{1}{2}\gamma}
    \\
    \frac{t_1^\frac{1}{2} -t_2^\frac{1}{2}\gamma}{1 - (t_1 t_2)\frac{1}{2}\gamma}
    \end{pmatrix}.
\end{equation}

Both $a^D$ and $a^T$ match the Feynman path integral results, as well as previously reported ones \cite{Rabus2007}.

\subsection{Quantum vs Classical Random Walk: Time Domain} \label{sec:timedomain}

\begin{figure}[b]
    \centering
    \includegraphics[width=9.5cm]{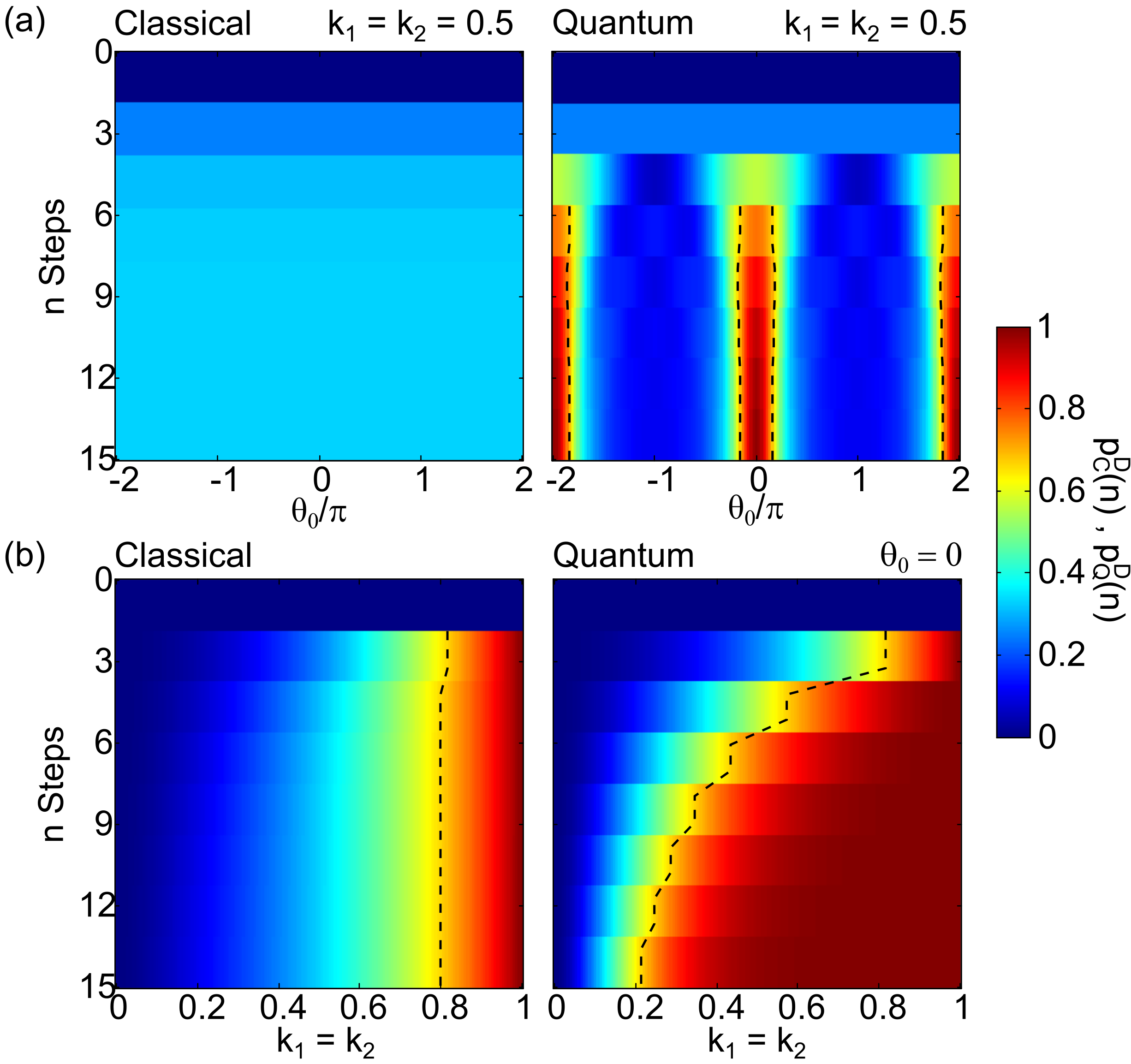}
    \caption{
    Classical and quantum
    random walk \textit{Drop} port probabilities of one-ring configuration as a function of time, for variable coherent phase $\theta$ (a) and hopping probabilities (coupling coefficients) $k_1 = k_2$ ($k_1 = k_2$) (b).
    The black dashed lines indicate the goal-hitting threshold $p_g = 2/3$.
    All plots share the same intensity color scale.
    }
    \label{fig:S:TimeDomain}
\end{figure}

Fig.~\ref{fig:S:TimeDomain} plots the random walk step-dependent probabilities of reaching the \textit{Drop} port for the classical and quantum random walks for the one-ring configuration.
These data were used for the comparison presented in Fig.~\ref{fig:QRW-CRW:TimeDomain} of the main text.
Figure~\ref{fig:S:TimeDomain}(a) presents a sweep over the coherent phase (constant in the classical case), while Fig.~\ref{fig:S:TimeDomain} sweeps over the matched hopping probabilities $k_1=k_2$ and coupling coefficients $k_1 = k_2$ for the classical and quantum walks, respectively.

\clearpage

\section{Random Walk on a Double Ring Graph}
\label{ap:RW:2rings}

\begin{figure}[b]
    \includegraphics[width=8.1cm]{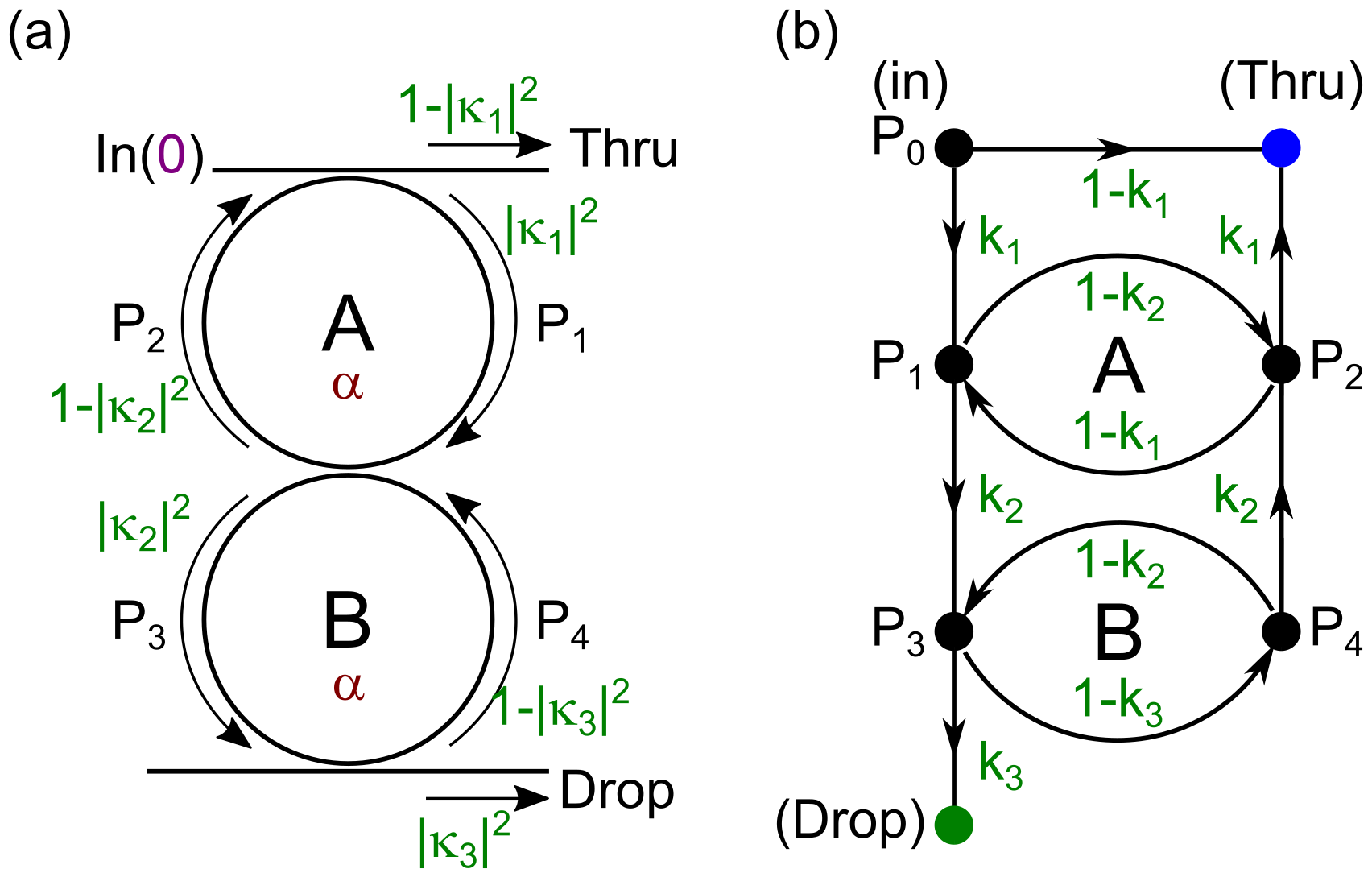}
    \caption{\label{fig:S:concept:2rings}
    Random walk on two coupled ring resonators. 
    (a) Implementation using a series-coupled double-ring resonator.
    $\kappa_i, i = \{1,2,3\}$ and $\alpha$ are the coupling and loss coefficients, respectively.
    (b) Graph used to model the random walk on the two-ring structure. $k_i = |\kappa_i|^2, i = \{1,2,3\}$ are the walker jump probabilities associated with each pair of nodes. The loss coefficient $\alpha$ is omitted in the figure but considered in the calculations.
} \label{fig:2rings}
\end{figure}

This section presents a complete description of the classical and quantum random walk on a series-coupled double photonic ring resonator structure.
Fig.~\ref{fig:S:concept:2rings} schematizes the double ring graph and associated hopping probabilities and coupling coefficients.
Similarly to the work presented in the \textit{main text}, the calculations are performed in two ways: 1) using a Feynman path sum; 2) using a Markov Chain approach.

\subsection{Classical Random Walk}
\label{ap:CRW:2rings}

Here we calculate the probabilities $p_C^D$ and $p_C^T$ for the classical random walk on a double ring graph, describing in detail the Feynman path sum and the Markov chain method.

\subsubsection{Feynman path sum}

The probability $p_C^D$ can be expressed (Fig.~\ref{fig:S:concept:2rings}(b)) as

\begin{equation}
    \label{eq:CRW:p0-pd}
    p_C^D = k_1 p_C(P_0 \to P_3) k_3 .
\end{equation}

Once the walker enters a ring, it can loop inside it, performing between $0$ to infinite full round-trips before coupling to another waveguide.
Let $p_A$ and $p_B$ be the probabilities $p_C(P_1 \to P_1)$ and $p_C(P_3 \to P_3)$ considering all infinite possible loops around the rings A and B, respectively.

\begin{align}
    p_A = \sum_{m=0}^{+\infty} (t_1 t_2 \alpha)^m = \frac{1}{1-t_1 t_2 \alpha}
&,& |t_1|,|t_2|,|\alpha| \leqslant 1
\\
    p_A = \sum_{n=0}^{+\infty} (t_2 t_3 \alpha)^n = \frac{1}{1-t_2 t_3 \alpha}
&,& |t_3| \leqslant 1
\end{align}

The probability $p_C(P_1 \to P_3)$ can be calculated by considering all possible ring round trips as follows 

\begin{eqnarray}
    p_C(P_1 \to P_3) = p_A p_B k_2 \alpha \sum_{s=0}^{+ \infty} (p_A)^s(p_B)^s(k_2^2 t_1 t_3 \alpha_2)^s \nonumber
\\
    = \frac{k_2 \alpha}{(1 - t_1 t_2 \alpha)(1 - t_2 t_3 \alpha) - k_2^2 t_1 t_3 \alpha^2},
\end{eqnarray}

\noindent where the term in evidence accounts for the minimum pathway between $P_1$ and $P_3$, while the sum accounts for all possible inter-ring transfers and loops.
Replacing $p_C(P_1 \to P_3)$ in Eq.~(\ref{eq:CRW:p0-pd}) yields

\begin{equation}
    \label{eq:ap:CRW:p0-pd-final}
    p_C^D 
    = \frac{k_1 k_2 k_3 \alpha}{1 - t_1 t_2 \alpha - t_2 t_3 \alpha - t_1 t_3 \alpha^2 + 2 t_1 t_2 t_3 \alpha^2}
\end{equation}

A similar approach can be used to calculate $p_C^T$, except that in this case, the walker can move directly to $P_T$ with a probability $t_1$

\begin{equation}
    \label{eq:CRW:p0-pt}
    p_C^T = t_1 + k_1^2 p_C(P_1 \to P_2)
\end{equation}

To calculate $p_C(P_1 \to P_2)$, we must consider two possibilities: 1) the walker loops around ring $A$ without visiting ring $B$, $p_C(P_1 \to P_2)_A$; and 2) the walker visits and loops around both rings $p_C(P_1 \to P_2)_{AB}$, so that

\begin{equation}
    p_C(P_1 \to P_2) = p_C(P_1 \to P_2)_A + p_C(P_1 \to P_2)_{AB}
\end{equation}

For a walker looping around ring $A$ only,

\begin{equation}
    p_C(P_1 \to P_2)_A = t_2 \alpha \sum_{m=0}^{+\infty} (t_1 t_2 \alpha)^m = \frac{t_2 \alpha}{1 - t_1 t_2 \alpha}.
\end{equation}

For a walker looping around both rings $A$ and $B$,

\begin{equation}
    p_C(P_1 \to P_2)_{AB} = 
    p_A \alpha^\frac{1}{2} \sum_{s=1}^{+\infty} (p_A p_B)^s (k_2^2 t_3 \alpha^{3/2})^s(t_1 \alpha^\frac{1}{2})^{s-1},
\end{equation}

\noindent where the sum must start in $s = 1$.
The term $p_A \alpha^\frac{1}{2}$ in evidence accounts for the infinite $A$ loops before reaching $B$, and the minimum pathway loss $\alpha^\frac{1}{2}$ in the first half of ring $A$ to reach $B$.
The term $(t_1 \alpha^\frac{1}{2})^{s-1}$ accounts for half-turns around ring $A$, which only occur for the second round trip on (hence the $s-1$ exponent).
Summing both possibilities and replacing in Eq.~(\ref{eq:CRW:p0-pt}) yields

\begin{align}
\label{eq:ap:CRW:p0-pt-final}
    p_C^T &= t_1 + 
    \frac{k_1^2 \alpha}{1 - t_1 t_2 \alpha} \left( t_2 + \frac{k_2^2 t_3 \alpha}{(1 - t_1 t_2 \alpha)(1 - t_2 t_3 \alpha) - k_2^2 t_1 t_3 \alpha^2} \right)\nonumber
    \\
    & = \frac{t_1 + t_2 \alpha - 2 t_1 t_2 \alpha - (t_1 t_2 - (1 - 2 t_1) (1 - 2 t_2)\alpha) t_3 \alpha}
    {1 - t_1 t_2 \alpha - t_2 t_3 \alpha - t_1 t_3 \alpha^2 + 2 t_1 t_2 t_3 \alpha^2}
\end{align}

Naturally, $p_C^D + p_C^T = 1$ for the lossless case of $\alpha = 1$.

\subsubsection{Markov Chain approach}
\label{ap:CRW:2rings:Markov}

Expanding on the formulation used for the single ring graph, the classical random walk on a double ring graph using the Markov Chain method can be expressed as

\begin{equation}
    p_C^{(n)} = \begin{pmatrix} p^0\\p^1\\p^2\\p^3\\p^4\\p^D\\p^T \end{pmatrix}_C^{(n)}
    = T^n \begin{pmatrix} 1\\0\\0\\0\\0\\0\\0 \end{pmatrix},
        T^1 =
    \begin{pmatrix}
0	&	0	            &	0	            &	0	            &	0	            &	0	&	0	\\
k1	&	0	            &	t_1\alpha^\frac{1}{2}&	0	            &	0	            &	0	&	0	\\
0	&	t_2\alpha^\frac{1}{2}&	0	            &	0	            &	k_2\alpha^\frac{1}{2}&	0	&	0	\\
0	&	k_2\alpha^\frac{1}{2}&	0	            &	0	            &	t_2\alpha^\frac{1}{2}&	0	&	0	\\
0	&	0	            &	0	            &	t_3\alpha^\frac{1}{2}&	0	            &	0	&	0	\\
0	&	0	            &	0	            &	k_3\alpha^\frac{1}{2}&	0	            &	1	&	0	\\
t1	&	0	            &	k_1\alpha^\frac{1}{2}	                &	0	            &	0	&	0	&	1	
	
    \end{pmatrix}
    \label{eq:Markcov:quantum:P}
\end{equation}

Calculating the eigenmatrix $P$, we obtain the node occupation probabilities as the number of steps goes to infinity as

\begin{equation}
    p^{\infty} = \left(P \left(P^{-1} T^1 P \right)^{\infty} P^{-1}\right) p^1 = 
    \begin{pmatrix}
    0\\0\\0\\0\\0\\
    \frac{k_1 k_2 k_3 \alpha}{1 - t_1 t_2 \alpha - t_2 t_3 \alpha - t_1 t_3 \alpha^2 + 2 t_1 t_2 t_3 \alpha^2}
    \\
    \frac{t_1 + t_2 \alpha - 2 t_1 t_2 \alpha - (t_1 t_2 - (1 - 2 t_1) (1 - 2 t_2)\alpha) t_3 \alpha}
    {1 - t_1 t_2 \alpha - t_2 t_3 \alpha - t_1 t_3 \alpha^2 + 2 t_1 t_2 t_3 \alpha^2}
    \end{pmatrix},
\end{equation}
a solution that matches $p_C^D$ and $p_C^T$ obtained via the detailed reasoning using Feynman path sums, as given by Eqs.~\ref{eq:ap:CRW:p0-pd-final} and \ref{eq:ap:CRW:p0-pt-final}.

\subsubsection{Probability Distribution analysis}

\begin{figure}[b]
    \centering
    \includegraphics[width=8.6cm]{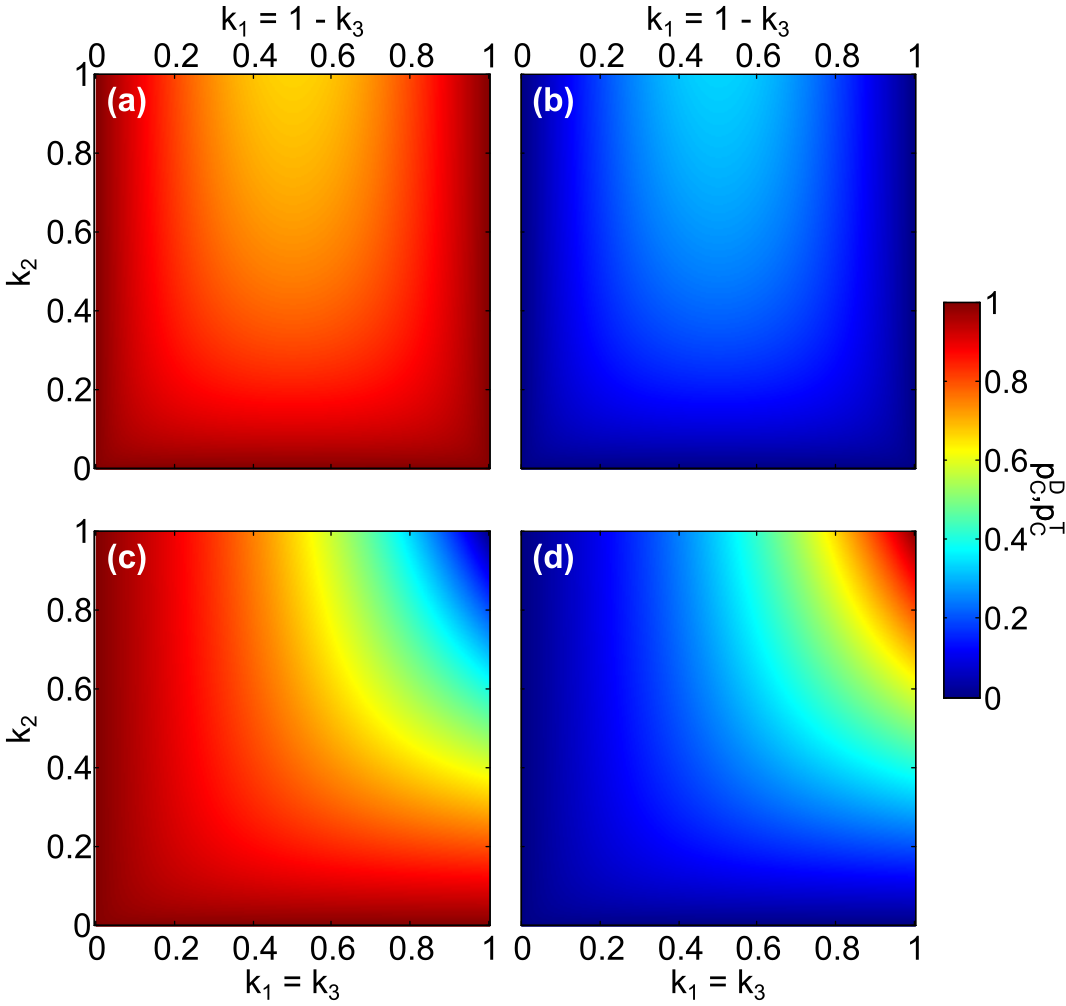}
    \caption{\textit{Thru} ($p_C^T$) and \textit{Drop} ($p_C^D$) distributions for the classical random walk on the double-ring graph of Fig. \ref{fig:2rings}-b.
    (a,b) Classical probability of a walker starting in $P_0$ and reaching $P_T$ (a) and $P_D$ (b), sweeping over the node hopping probabilities $k_1$ and $k_2$, with $k_3 = 1 - k_1$.
    (c,d) Analogous to (a,b), except $k_3 = k_1$.}
    \label{fig:S:CRW-2Rings}
\end{figure}

Fig.~\ref{fig:S:CRW-2Rings} shows the $p_C^T$ and $p_C^D$ distributions for the classical random walk on a double-ring graph, resulting for different $k_i$ combinations.
Due to the graph symmetry, we decide to link $k_1$ and $k_3$ and focus on two contrasting scenarios: \textit{Scenario 1}, $k_3 = 1 - k_1$; and \textit{Scenario 2}, $k_3 = k_1$, which can be seen as different puzzles.
In \textit{Scenario 1}, the probability of the walker hopping to the \textit{Thru} port in the first step ($t_1 = 1 - k_1$) is equal to the probability of it hopping to the \textit{Drop} port ($k_3$).
However, once the walker enters the first ring, the probability of hopping to the \textit{Thru} port changes to $k_1$, and the puzzle becomes imbalanced.
Conversely, in \textit{Scenario 2}, the hopping probabilities towards the \textit{Thru} and \textit{Drop} ports are imbalanced in the first step but balanced throughout the rest of the walk.
The probability distributions associated with \textit{Scenario 1} and \textit{Scenario 2} are represented in Fig.~\ref{fig:S:CRW-2Rings}(a,b) and Fig.~\ref{fig:S:CRW-2Rings}(c,d), respectively, for sweeping combinations of $k_1$ and $k_2$.
Notice that $k_3 = t_1 \Longleftrightarrow k_1 = 1-k_3 = t_3$.

Let us consider a maze-like problem, where we want to maximize propagation to the \textit{Drop} port, i.e., maximize $p_C^D$.
Fig.~\ref{fig:S:CRW-2Rings}(b,d) shows that \textit{Scenario 1} is much less efficient than \textit{Scenario 2}, as it can only achieve a maximum efficiency of $1/3$ for $k_1 = t_3 = \frac{1}{2}$ and $k_2 = 1$.
In contrast, \textit{Scenario 2} can reach an efficiency all the way up to $1$.

\subsection{Quantum Random Walk}
\label{ap:QRW:2rings}

Appendix~\ref{ap:CRW:2rings:Markov} reports on the Feynman path sum and Markov chain methods to calculate the classical random walk \textit{Thru} and \textit{Drop} probabilities for the two-ring configuration.
Since the quantum random walk calculations are analogous to the classical ones, this section presents only the more straightforward Markov chain approach.


Here we calculate probability amplitudes corresponding to \textit{Drop} and \textit{Thru} ports for a series- coupled double-ring resoonator, using the Markov Chain approach. The transition amplitudes can be collected in a transition matrix $T^1$, making reference to the vertex ordering of Fig. \ref{fig:2rings}-b:

\begin{equation}
    T^1 =
    \begin{pmatrix}
0	                &	0	                        &	0	                        &	0	                        &	0	                        &	0	&	0	\\
-k_1^\frac{1}{2}    &	0	                        &	(t_1\gamma_1)^\frac{1}{2}   &	0	                        &	0	                        &	0	&	0	\\
0	                &	(t_2\gamma_1)^\frac{1}{2}   &	0	                        &	0	                        &	-(k_2\gamma_2)^\frac{1}{2}  &	0	&	0	\\
0	                &	(k_2\gamma_1)^\frac{1}{2}   &	0	                        &	0	                        &	(t_2\gamma_2)^\frac{1}{2}      &	0	&	0	\\
0	                &	0	                        &	0	                        &	(t_3\gamma_2)^\frac{1}{2}   &	0	                        &	0	&	0	\\
0	                &	0	                        &	0	                        &	(k_3\gamma_2)^\frac{1}{2}   &	0	                        &	1	&	0	\\
t_1^\frac{1}{2}	    &	0	                        &	(k_1\gamma_1)^\frac{1}{2}   &	0	                        &	0	                        &	0	&	1	

    \end{pmatrix}
\end{equation}

Finally, the steady-state occupation probabilities corresponding to each node can be found using Eq.~(\ref{eq:Markov:pNmain}) as

\begin{equation}
    a^{(\infty)} = \begin{pmatrix} a^0\\a^1\\a^2\\a^3\\a^4\\a^D\\a^T \end{pmatrix}^{(\infty)}
    = \left(P \left(P^{-1} T^1 P \right)^{ \infty} P^{-1}\right)
    \begin{pmatrix} 1\\0\\0\\0\\0\\0\\0 \end{pmatrix}
    = 
    \begin{pmatrix}
    0\\0\\0\\0\\0\\
    -\frac{(k_1 k_2 k_3 \gamma_1\gamma_2)^\frac{1}{2}}
    {1 - (t_3 t_2)^\frac{1}{2} \gamma_2 - (t_2 t_1)^\frac{1}{2} \gamma_1 + (t_3 t_1)^\frac{1}{2} \gamma_1 \gamma_2}
    \\
    \frac{t_1^\frac{1}{2} - t_2^\frac{1}{2} \gamma_1 - (t_1 t_2 t_3)^\frac{1}{2} \gamma_1 + t_3^\frac{1}{2} \gamma_1 \gamma_2}
    {1 - (t_3 t_2)^\frac{1}{2} \gamma_2 - (t_2 t_1)^\frac{1}{2} \gamma_1 + (t_3 t_1)^\frac{1}{2} \gamma_1 \gamma_2}
    \end{pmatrix}
\end{equation}

\begin{figure}
    \includegraphics[width=8.3cm]{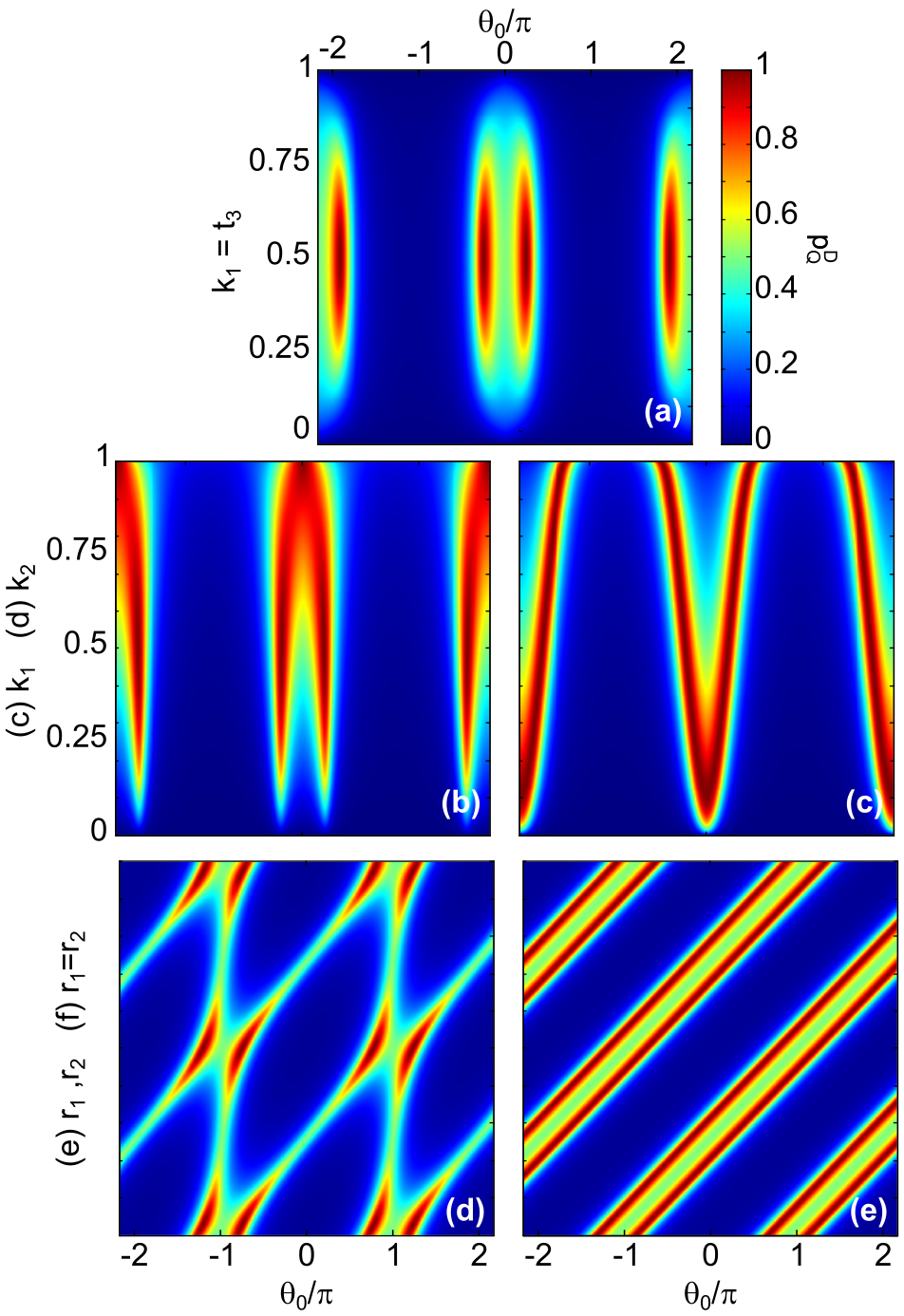}
    \caption{Quantum random walk in a series-coupled double ring resonator.
    (a-e) \acrshort{rr} \textit{Drop} port probability modulated by phase $\theta_0$ (wavelength) and 
    (a) variable coupling coefficients $k_1 = t_3$, while keeping $k_2^\frac{1}{2}$ constant;
    (b) variable $k_1$, while keeping $k_2,k_3$ constant;
    (c) variable $k_2$, while keeping $k_1,k_3$ constant;
    (d) variable ring radii $r_1$ or $r_2$, while keeping the other constant;
    (f) variable $r_1 = r_2$;
    When kept constant, $r_i = 20$~\micron{} and $k_i = 1/2$.
    All plots share the same color bar.
    }
    \label{fig:S:QRW:PD-2Rings}
\end{figure}

Fig.~\ref{fig:S:QRW:PD-2Rings} shows the strong dependence of $p_Q^D$ on the resonant constructive and destructive interference effects modulated by the coupling coefficients (Fig.~\ref{fig:S:QRW:PD-2Rings}(a-c)) and ring radii (Fig.~\ref{fig:S:QRW:PD-2Rings}(d,e)).

Double-peaked periodic resonances ($\theta_i/(2\pi) = n_{eff}L_i/\lambda = n, n \in \mathbb{N}$) are observed for any fixed set of $k_1^\frac{1}{2},k_2^\frac{1}{2},r_1,r_2$.
Changing $k_1^\frac{1}{2}$ and $k_2^\frac{1}{2}$ simultaneously and holding $k_3 = 1-k_1 = t_1$ modulates the resonance peak contrast, Fig.~\ref{fig:S:QRW:PD-2Rings}(a).
Meanwhile, increasing $k_1$ ($k_2$) individually (holding $k_3 = \frac{1}{2}$) increases (decreases) $p_Q^D$ within the spectral gap between resonances, Fig.~\ref{fig:S:QRW:PD-2Rings}(b,c).  
Changing either $r_1$ or $r_2$ individually modifies the spectral gap between resonance peaks, Fig.~\ref{fig:S:QRW:PD-2Rings}(d), and changing both radii simultaneously shifts the resonances, Fig.~\ref{fig:S:QRW:PD-2Rings}(e).

\clearpage

\section{Feasibility studies}
\label{ap:QRW:Feasibility}

Polymeric waveguides were fabricated on Si substrates by spin-coating a $20$~\micron{}-thick film of Epoclad cured by UV and hard-bake in oven, followed by the spin-coating of a $2.2$~\micron{} thick EpoCore layer.
The EpoCore layer is patterned by e-beam lithography followed by a development.
A hard bake is performed after the development in the oven to ensure the chip's thermal stability.

The light coupling coefficient $\kappa^2$ in a directional waveguide coupler depends on the ratio between the coupler length and the \textit{beat length} as follows:

\begin{align}
    L_b = & \frac{1}{2}\frac{\lambda}{n_{eff1}-n_{eff2}}
    \label{eq:ap-exp-Coupler-Lb}
    \\
    L_e = & L_s + 2r_b\mathrm{cos}^{-1}\left(1-\frac{d_c-(d-2r_w)}{2r_b+2r_w}\right)
    \label{eq:ap-exp-Coupler-Le}
    \\
    \kappa^2 = & \mathrm{sin}^2\left(\frac{\pi}{2}\frac{L_e}{L_b}\right)
    \label{eq:ap-exp-Coupler-k}
\end{align}

\noindent where $\lambda$ is the wavelength, $n_{eff1}$ and $n_{eff2}$ are the symmetric and asymmetric effective refractive indices, $d$ and $L_s$ are the coupler distance and length, $d_c$ is the minimum coupling distance, $r_w$ is the ridge half-width, $r_b$ is the bend radius, $L_b$ is the beat length, $L_e$ is the effective coupler length, and $\kappa^2$ is the coupling coefficient.

The effective refractive indices $n_{eff1}$ and $n_{eff2}$ are calculated via numerical mode solving \cite{Fallahkhair2008}.
The effective coupler length $L_e$ accounts for the partial coupling that occurs in the curved part of the waveguides, as they approach and depart from the straight coupler segment.

The bending losses in planar polymeric waveguides were characterized using a micro-lensed optical fiber (LFM1F-1, Thorlabs) side-coupling system (7TF2, Standa) and measuring the top-view scattering using a CMOS camera (CS2100M-USB, Thorlabs), with a 4x objective (PLN 4x/0.1, Olympus) and an $f=3.5$~cm lens.
The camera is translated by a manual translation stage (PT3A/M, Thorlabs).

\begin{figure}[b]
    \centering
    \includegraphics[width=13cm]{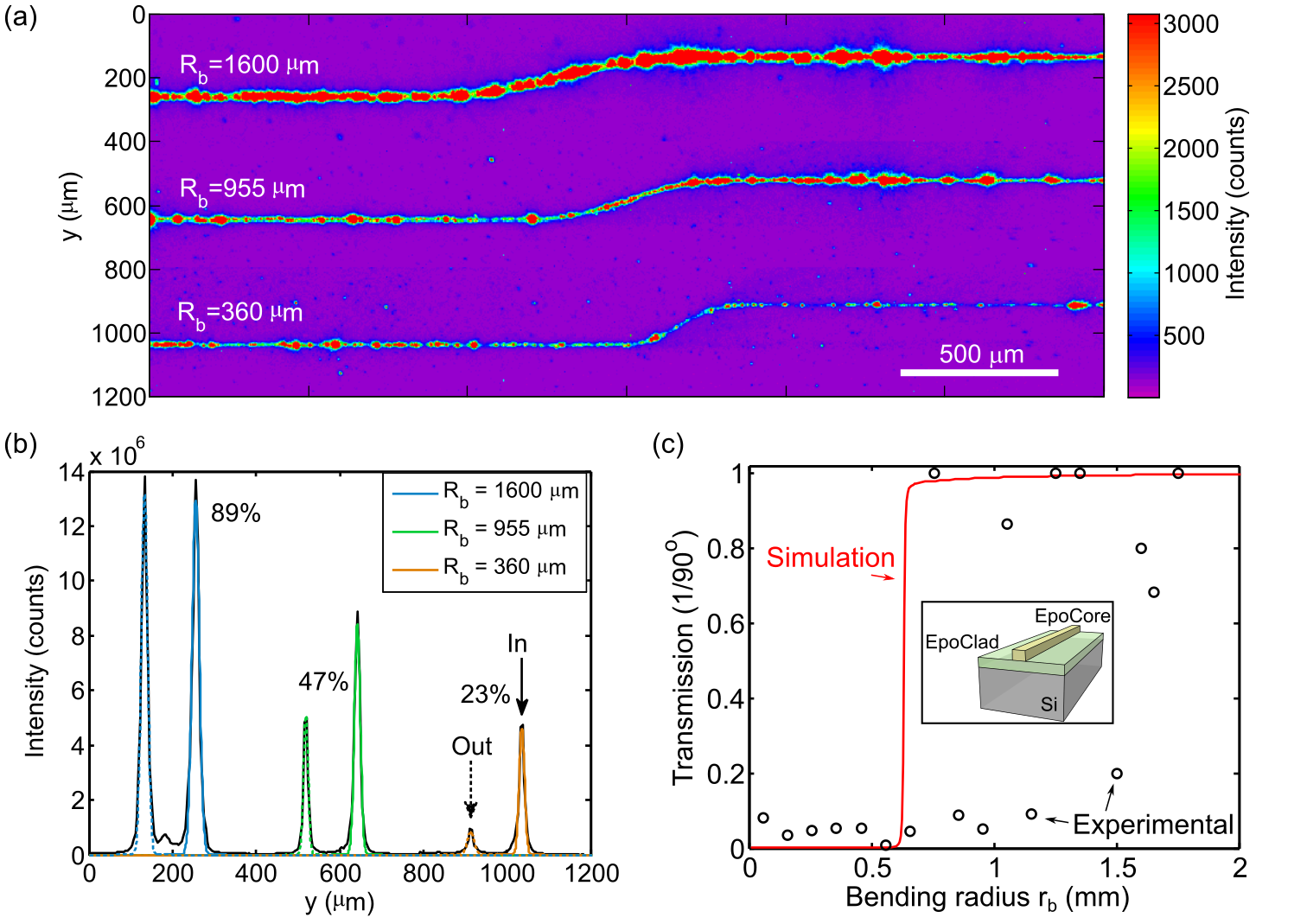}
    \caption{Experimental characterization of bending losses in planar polymeric waveguides.
    (a) Top-view light-scattering intensity of optical fiber-coupled polymeric planar bent waveguides with bending radii $r_b = {360, 955, 1600}$~\micron{}.
    (b) Horizontally-integrated intensities and eye-guiding Gaussian curve-fittings.
    The left and right peaks in each peak pair correspond to the output and input, respectively.
    (c) Image area- and curvature angle-normalized transmission per $90^{\circ}$ as a function of the bending radius $r_b$.
    The simulated transmission is plotted in red, while the measured values are plotted as circles.
    The inset figure illustrates the waveguide structures.
    Excitation wavelength $\lambda=635$~nm. 
    }
    \label{fig:S:exp-BendLosses}
\end{figure}

Fig.~\ref{fig:S:exp-BendLosses}(a) shows three examples of illuminated bent waveguides with bending radii of $360$, $955$, and $1600$~\micron{}.
Before and after the bends, the intensities are characterized using a MATLAB script that integrates the intensity images in the horizontal direction.
Fig.~\ref{fig:S:exp-BendLosses}(b) shows the integrated intensities and eye-guiding Gaussian fittings for bending radii of $360$, $955$, and $1600$~\micron{}.
The integrated intensities are normalized by the collection area curvature angle (both differ from image to image) and plotted as the transmission per $90^{\circ}$ in Fig.~\ref{fig:S:exp-BendLosses}(c) as a function of the bending radius $r_b$.
The simulation (Mode, Lumerical) of the same quantity is overlaid with the experimental results, showing a sharp transmission decrease for $r_b < 0.6$~mm.
The experimental results display a partial agreement with this prediction (very large transmission for a few data points, $r_b > 0.6$~mm), but with a considerable variance (very low transmission for a few data points, $r_b > 0.6$~mm).

The large variance in the experimentally obtained transmissions can be partly explained by the poor data statistics (single measurement per $r_b$) and the characterization method employed (top-view characterization means optimizing the scattering a not necessarily waveguide transmission).
Nanofabrication uncertainty such as possible e-beam exposure errors, dust particles, and roughness of the diced waveguide inputs/outputs also contribute greatly to the experimental uncertainty.

\end{document}